\documentclass{article}
\pdfoutput=1

\usepackage[utf8]{inputenc}
\usepackage[style=apa,sortcites=true,sorting=nyt,backend=biber]{biblatex}
\usepackage{amsmath}
\usepackage{nicefrac}
\usepackage{mathbbol}
\usepackage{mathtools}
\usepackage{subcaption}
\usepackage{graphicx}
\usepackage{pdfpages,caption}
\usepackage{chngcntr}
\usepackage{apptools}
\newtheorem{theorem}{Theorem}
\newtheorem{lemma}[theorem]{Lemma}

\AtAppendix{\counterwithin{theorem}{section}}
\usepackage{appendix}
\usepackage{fancyvrb}
\usepackage{hyperref}
\usepackage{titlesec}
\usepackage{rotating}
\usepackage{float}
\usepackage{enumitem}
\usepackage{listings}
\usepackage{xcolor}
\usepackage{afterpage}
\usepackage{comment}
\usepackage{cancel}
\usepackage{booktabs}

\usepackage{arxiv}

\titleformat{\paragraph}
{\normalfont\normalsize\bfseries}{\theparagraph}{1em}{}
\titlespacing*{\paragraph}
{0pt}{3.25ex plus 1ex minus .2ex}{1.5ex plus .2ex}
\DefineVerbatimEnvironment{code}{Verbatim}{fontsize=\small, frame=single}

\title{Bridging Item Response Theory and Factor Analysis: \\
A Four-Parameter Mixture-Dichotomized Model with Bayesian Estimation}

\renewcommand{\shorttitle}{
Bridging Item Response Theory and Factor Analysis: 4P FA Model with Bayesian Estimation }

\author{
Ján Pavlech$^{1}$, Patrícia Martinková$^{1,2}$\\
\small $^{1}$ Institute of Computer Science of the Czech Academy of Sciences, Prague, Czech Republic\\
\small $^{2}$ Faculty of Education, Charles University, Prague, Czech Republic\\
}
\date{}
\begin{document}

\maketitle
\begin{abstract}
Item Response Theory (IRT) and Factor Analysis (FA) are two major frameworks for modeling multi-item measurements of latent traits. While the relationship between two-parameter IRT models and dichotomized FA models is well established, 
FA formulations for IRT models with additional parameters are less common. %, and relationships and computational comparisons between the two modeling frameworks are not sufficiently explored. 
We focus on the four-parameter factor-analytic (4P FA) model that extends the traditional dichotomized single-factor FA model through a hierarchical mixture formulation accounting for guessing and inattention effects. We analytically establish the equivalence of the 4P FA and 4P IRT models, extending the FA--IRT correspondence beyond the two-parameter case. A Bayesian estimation procedure is developed for model estimation, 
to estimate the four item parameters, the respondents' latent scores, and the scores adjusted for guessing and inattention effects. The proposed algorithm is implemented in \texttt{R} and \texttt{Python}. A simulation study compares estimation under the FA and IRT formulations of the 4P model and evaluates the practical implications of the FA parametrization. Empirical examples based on an admission test and an anxiety inventory demonstrate the correspondence between the 4P FA and 4P IRT models and illustrate the application of the proposed methodology.
\end{abstract}
\newpage

%\pagenumbering{arabic} 
% keywords: item response theory, factor analysis, mixture model, Bayesian estimation

%----------------------------------------------------------------------------------------------
\section{Introduction}
Item Response Theory (IRT) and Factor Analysis (FA) are foundational psychometric frameworks used to model multi-item data capturing unobserved psychological constructs, such as intelligence, cognitive ability, anxiety, or personality traits. Both frameworks rely on latent variables to represent respondent traits and use item-specific parameters to capture item functioning ~\parencite{rao2007psychometrics,bartholomew2008analysis,martinkova2023computational}. 
In the case of binary items, the relationship between the two-parameter (2P) IRT model and the dichotomized one-factor FA model is well established. Analytical transformations allow for direct mapping between IRT discrimination/difficulty parameters and FA loadings/thresholds \parencite{lord2008statistical, takane1987relationship, McDonald1999, kamata2008note}. This correspondence has played an important role in connecting the IRT and factor-analytic traditions of latent variable modeling, demonstrating that models originating from different psychometric frameworks can often be viewed as alternative parameterizations of the same underlying structure.

Extensions in the IRT framework, such as the 3-parameter and 4-parameter models (3P and 4P IRT), incorporate pseudo-guessing and inattention parameters to account for response behaviors like guessing or careless mistakes in cognitive tests, or pretending and dissimulation in psychological assessments \parencite{barton1981upper}. These models have been the focus of active methodological research 
\parencite[][]{battauz2020regularized, culpepper2016revisiting, loken2010estimation, meng2020marginalized, fu2021gibbs, waller2017bayesian}.
While the equivalence between the 2P IRT model and dichotomized FA models is well established, analogous FA formulations for 3P and 4P IRT models are less common. 
Although the parameter mapping underlying the 2P FA--IRT equivalence naturally suggests corresponding relationships for discrimination and difficulty parameters, extending the equivalence to 3P and 4P models is not straightforward because the guessing and inattention parameters introduce lower and upper asymptotes that cannot be obtained from the standard dichotomized FA model and therefore require a richer latent-variable formulation. Additionally, alternative model specifications and estimation frameworks may offer practical advantages, particularly for higher-parameter models, where estimation is computationally demanding and can be sensitive to convergence and identifiability issues. 

To address this gap, we focus on a three- and four-parameter extensions of the FA model (3P FA and 4P FA). Specifically, we develop factor-analytic counterparts of 3P and 4P IRT models by extending the dichotomized FA framework with a hierarchical mixture structure that accounts for guessing and inattention effects. 
%while we also discuss the implementation challenges of conventional covariance-based approaches. 
In addition to estimating item parameters, the model provides respondent ability estimates adjusted for guessing and inattention effects. 

The contributions of this work are threefold. First, we introduce a hierarchical mixture extension of the dichotomized FA model for binary items and formally establish its equivalence to the corresponding 4P IRT model, extending the FA--IRT correspondence to models with guessing and inattention parameters. Second, we develop 
a Bayesian estimation procedure for the proposed model and provide open-source implementations in both \texttt{R} and \texttt{Python}. Third, we conduct a simulation study that compares estimation under the FA and IRT formulations of the 4P model and demonstrate that the FA parameterization can offer practical advantages in estimation performance.

The remainder of the paper is structured as follows. Section~\ref{methods} begins by revisiting the 4P IRT model along with its restricted 3P and 2P versions. We then review the dichotomized single-factor FA model and its established connection to the 2P IRT model. Building on this foundation, we introduce the 4P FA model as a counterpart of the 4P IRT model, with formal derivations of the equivalence deferred to Appendix~\ref{appendix}. We also present a Bayesian estimation procedure for the 4P FA model and describe its implementation in both \texttt{R} and \texttt{Python}. In Section~\ref{simstudy}, a simulation study is conducted to compare the performance of the proposed Bayesian estimation method and alternative Bayesian and EM approaches using the IRT parametrization. Section~\ref{realdata} provides two empirical examples that illustrate the application of the proposed model and demonstrate the correspondence between the 4P FA and 4P IRT formulations of the model. Finally, Section~\ref{discussion} offers a discussion of the findings, implications for practice, and directions for future work. 

%----------------------------------------------------------------------------------------------

\section{Methods}\label{methods}

%-----------------
\subsection{The four-parameter IRT model}\label{4P_IRT}
We begin by defining the 2-parameter logistic (2PL) and 2-parameter normal ogive (2PNO) IRT models. We assume that the latent trait $\theta$ follows a continuous standardized distribution with density $f(\theta)$, such as the standard normal distribution $\mathsf{N}(0, 1)$. 
The item response curve (IRC) defines the probability that binary response $Y_i$ on item $i$ equals one, conditioned on the ability~$\theta$, and takes the form:

\begin{align}\label{eq:2P-IRT}
    \textsf{P}(Y_{i}=1| \theta)= F\big(a_{i}(\theta-b_{i})\big),
\end{align}
for some reasonable (non-decreasing, continuous) function $F$. The most common choices of $F()$ are the cumulative distribution function of the standard normal distribution, yielding the 2PNO IRT model, or the logistic function $\frac {e^x}{1 + e^x}$, yielding the 2PL IRT model \parencite{birnbaum1968statistical, lord2008statistical}. The interpretation of item parameters $a_i$ and $b_i$ varies by context. In educational testing, $b_i$ represents \textit{item difficulty}, that is, the ability level at which a respondent has a 50\% probability of answering item $i$ correctly. In psychological or health-related assessment, $b_i$ is sometimes called \textit{item popularity}. In both settings, $a_i$ is referred to as the \textit{item discrimination} parameter, indicating the sensitivity of the probability of a correct or endorsed response to changes in $\theta$. 

\vspace{1em}

Two additional parameters can be incorporated into Equation~\eqref{eq:2P-IRT}, yielding the 4P IRT model \parencite{barton1981upper}: 
\begin{align}\label{eq:4P-IRT}
    \textsf{P}(Y_{i}=1| \theta)=c_i + (d_i - c_i) \cdot F\big(a_{i}(\theta-b_{i})\big).
\end{align}

The lower left asymptote $c_i \in [0,1)$ is referred to as the \textit{pseudo-guessing} parameter in educational contexts, representing the probability of answering the item correctly despite lacking the respective knowledge---typically due to guessing. In psychological assessments, it may also be interpreted as a \textit{pretending} parameter. 

The upper asymptote $d_i \in (c_i,1]$ accounts for lapses such as \textit{inattention} or \textit{slipping} in educational settings, and \textit{dissimulation} in psychological contexts. 
For simplicity, we use the terminology aligned with educational testing---namely, ability, item difficulty, guessing, and inattention---unless psychological applications are explicitly discussed. 

As before, the choice of function $F()$ determines the specific variant of the model. The 4-parameter normal ogive (4PNO) IRT model uses the standard normal cumulative distribution, while the 4-parameter logistic (4PL) IRT model uses the logistic function $\frac {e^x}{1 + e^x}$, corresponding to the logistic distribution $logist(0,1)$.

The responses to individual items are assumed to be conditionally independent given $\theta$. Under this assumption, the probability of an entire response pattern $\mathbf{y} = (y_{1}, \dots, y_{m})^T$ is given by: 

\begin{align}\label{2}
    \textsf{P}(\mathbf{Y} = \mathbf{y}| \theta) = \prod_{i = 1}^m  \Big[\textsf{P}(Y_{i} = 1 | \theta)\Big]^{y_i} 
     \Big[1 - \textsf{P}(Y_{i} = 1 | \theta)\Big]^{1 - y_i}.
\end{align}

The 3P IRT model and the 2P IRT model \eqref{eq:2P-IRT} are both nested within the 4P model~\eqref{eq:4P-IRT}. The 3P model is obtained by setting $d_i = 1$ (i.e., no inattention), and the 2P model results from additionally setting $c_i = 0$  (i.e., no pseudo-guessing).

%-----------------
\subsection{Dichotomized 2P FA model and its relation to the 2P IRT model}
\label{sec:FA-IRT}
The linear FA model assumes a linear relationship between the vector of continuous 
item scores $\mathbf{Y}^*$ and a single latent variable $\theta$

\begin{align}\label{eq:FA1}
    \mathbf{Y}^* = \mathbb{A} \theta + \boldsymbol{\epsilon}.
\end{align}
 
Here, $\mathbf{Y}^*$ is a vector of length $m$ representing the item scores, $\mathbb{A}=(\alpha_{1}, \dots, \alpha_{m})^T$ is the vector of factor loadings, and $\theta$ is a standardized latent trait with density $f(\theta)$. The error vector $\boldsymbol{\epsilon}$ has independent components, each distributed as either normal $\mathsf{N}(0, u_{i}^2)$ or logistic $logist(0, u_{i}^2)$, with $u_{i}^2 = 1 - \alpha^2_{i}$ representing item  \textit{uniqueness}. Independence between $\theta$ and $\boldsymbol{\epsilon}$ is assumed.

For binary item responses, $\mathbf{Y}^*$ is not observed directly. Instead, observed binary item response, $Y_i$ result from dichotomizing $Y^*_i$ at an unknown threshold $\tau_i$, or equivalently, by regions $R_i$ as follows:
\begin{align}\label{eq:FA2}
    Y_{i} = \begin{cases}
        0, \; \; \; \; Y^*_{i} < \tau_{i} \; &||  \; \; R^0_{i} = (-\infty, \tau_{i}),\\
        1, \; \; \; \; Y^*_{i} \geq \tau_{i} \; &|| \; \; R^1_{i} = \hspace{0.08cm}[\tau_{i}\phantom{,,},\infty).
       \end{cases}
\end{align}
Equations  \eqref{eq:FA1}  \eqref{eq:FA2} define the dichotomized 2P FA model. This model has been shown to be equivalent to the 2P IRT model  \eqref{eq:2P-IRT} with parameter transformations given by 
\parencite[][]{takane1987relationship, kamata2008note}:

\begin{align}\label{eq:eqvi}
    &a_{i} = \frac{\alpha_{i}}{u_{i}} = \frac{\alpha_{i}}{\sqrt{1 - \alpha_{i}^2}}, \; \; \; b_{i} = \frac{\tau_{i}}{\alpha_{i}},\; \; \; or\nonumber\\
    \,\nonumber\\
    &\alpha_{i} = \frac{a_{i}}{\sqrt{1 + a_{i}^2}}, \; \; \; \tau_{i} = \frac{a_{i}b_{i}}{\sqrt{1 + a_{i}^2}},\; \; i = 1,\dots,m.
\end{align}
These equations relate IRT item discrimination $a_i$ and difficulty parameters to FA factor loadings $\alpha_i$ and thresholds $\tau_i$.
Alternative parameterizations and corresponding formulae are discussed by~\cite{kamata2008note}. 
When switching between logistic and normal link functions, the parameters can be approximately converted using a scaling factor of 1.7 \parencite{haley1952estimation}.

%-----------------
\subsection{The 3P and 4P model in factor analytic framework}\label{model_spec}
To construct a 3P FA model equivalent to the 3PL or 3PNO IRT model, we introduce a second probabilistic layer. Specifically, for respondents whose continuous latent item response $Y^*_{i}$ falls below the threshold~$\tau_{i}$, the observed response $Y_i$ is drawn from a Bernoulli distribution with success probability $c_i$, representing the guessing parameter. Conversely, for those with $Y^*_{i} \geq \tau_{i}$, a correct response is deterministically observed:

\begin{align}\label{eq:3P-FAv1}
    Y_{i} = \begin{cases}
        \sim Bernoulli(c_{i}), \; \; \; \; Y^*_{i} < \tau_{i} \; &||  \; \; R^0_{i} = (-\infty, \tau_{i}),\\
        1 \phantom{1Bernoulli(c_{i})}, \; \; \; \; Y^*_{i} \geq \tau_{i} \; &|| \; \; R^1_{i} = \hspace{0.08cm}[\tau_{i}\phantom{,,},\infty).
       \end{cases}
\end{align} 

Equivalently, and more formally, we assume a continuous latent item response $Y^*_{i}$ defined by~\eqref{eq:FA1} and~\eqref{eq:FA2}, then we define a \textit{discrete binary latent responses} $Z_{i}$  
\begin{align}\label{eq:Zi}
    Z_{i} = \begin{cases}
        0, \; \; \; \; Y^*_{i} < \tau_{i} \; &||  \; \; R^0_{i} = (-\infty, \tau_{i}),\\
        1, \; \; \; \; Y^*_{i} \geq \tau_{i} \; &|| \; \; R^1_{i} = \hspace{0.08cm}[\tau_{i}\phantom{,,},\infty).
       \end{cases}
\end{align}
 And finally, we consider the observed binary response $Y_i$ to be a mixture of a Bernoulli distribution and constant 1 based on the value of $Z_i$:
\begin{align}\label{eq:3P-FA}
    Y_{i} = \begin{cases}
        \sim Bernoulli(c_{i}), \; \; \; \; Z_{i} = 0,\\
        1 \phantom{1Bernoulli(c_{i})}, \; \; \; \; Z_{i} = 1.
       \end{cases}
\end{align}
Here, $c \in [0,1)$ models the probability of a correct response due to guessing. The interpretation is as follows: The respondent with latent ability $\theta$ has latent item performance $Y^*_{i}$. The first-level dichotomization~\eqref{eq:Zi} leading to latent binary variable $Z_i$ indicates whether the respondent was able to solve item $i$. More specifically, if their latent item performance $Y^*_{i}$ was higher than the threshold $\tau_{i}$, they were able to solve the item correctly, i.e., $Z_i = 1$. On the other hand, if their latent item performance was lower than $\tau_{i}$, they could not solve the item. The second level mixture~\eqref{eq:3P-FA} yielding the observed binary response $Y_i$ incorporates the information whether the respondent guessed or not as follows: If the respondent was not able to solve the item, then they may have guessed the answer, so there was still the probability $c_{i}$ of answering correctly. 
Note that for $c_{i} = 0,$ dichotomization \eqref{eq:3P-FA} reduces to \eqref{eq:FA2}, and the 3P FA model reduces to the dichotomized 2P FA model.

\vspace{1em}

To construct the 4P FA model equivalent to the 4PL or 4PNO IRT model, we extend this framework to incorporate both guessing and inattention effects by adding a second level mixture in~\eqref{eq:3P-FAv1} for the case when the respondent's continuous latent response $Y^*_{i}$ falls above the threshold $\tau_{i}$:

\begin{align*}
    Y_{i} = \begin{cases}
        \sim Bernoulli(c_{i}), \; \; \; \; Y^*_{i} < \tau_{i} \; &||  \; \; R^0_{i} = (-\infty, \tau_{i}),\\
        \sim Bernoulli(d_{i}), \; \; \; \; Y^*_{i} \geq \tau_{i} \; &|| \; \; R^1_{i} = \hspace{0.08cm}[\tau_{i}\phantom{,,},\infty).
       \end{cases}
\end{align*}

Equivalently, using a discrete latent response $Z_{i}$ as in~\eqref{eq:Zi}: 
\begin{align}\label{eq:4P-FA}
    Y_{i} = \begin{cases}
        \sim Bernoulli(c_{i}), \; \; \; \; Z_{i} = 0,\\
        \sim Bernoulli(d_{i}), \; \; \; \; Z_{i} = 1.
       \end{cases}
\end{align}
Here, $c_i \in [0,1)$ is the pseudo-guessing parameter and $d_i \in (0,1]$, $c_i < d_i$ is the inattention parameter. This structure accounts for both false positives (guessing) and false negatives (inattention).

The mixture-dichotomized model defined by equations \eqref{eq:FA1},~\eqref{eq:Zi}, and~\eqref{eq:4P-FA} is termed the 4P FA model. It is analytically equivalent to the 4P IRT model, as we show in Appendix~\ref{sec:formal}. The parameters $\alpha_{i}$ and $\tau_{i}$ of the 4P FA model map to $a_{i}$ and $b_{i}$ of the 4P IRT model through transformations~\eqref{eq:eqvi}, and the $c_{i}$ and $d_{i}$ are shared directly. Two model variants are defined depending on the error distribution in $\boldsymbol{\epsilon}$: if the elements follow a normal $\mathsf{N}(0, u_{i}^2)$ distribution, this leads to the 4PNO FA model, whereas if they follow the logistic $logist(0, u_{i}^2)$ distribution, this leads to 4PL FA model. The equivalence of the 4P FA model and the respective 4P IRT model is established by proving that the conditional probabilities defining the IRCs are equal for the two models (see Appendix~\ref{sec:formal:conditional}) and that the marginal probabilities of the response patterns also match (see Appendix~\ref{sec:formal:marginal}).
As a result, the 4P FA model is equivalent to any reparameterized form of the 4P IRT. 

%-----------------
\subsection{Estimation with Bayesian samplers}\label{implementation}

The 4P FA model is hierarchically structured, making it well-suited for parameter estimation within the Bayesian framework. Moreover, Bayesian estimation methods can be applied to more typical IRT parameterizations, which model the probability of a correct response directly.

\subsubsection{FA parameterization}

\begin{figure}[!ht]
    \centering
    \includegraphics[scale=0.7]{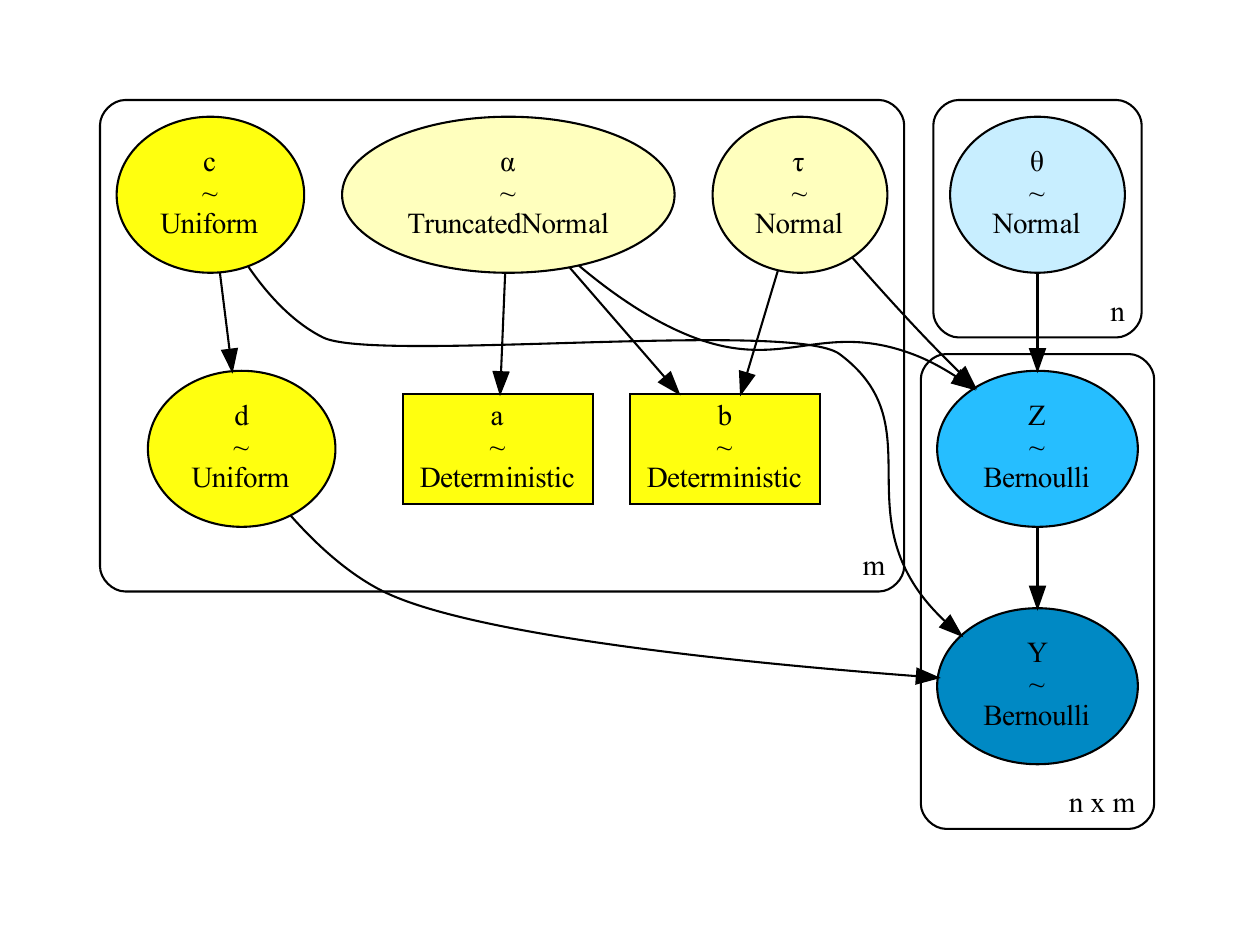}
     \vspace{-13mm}
    \caption{Directed Acyclic Graph (DAG) of hierarchically specified 4P FA model with m items and n respondents. }
    \label{figDAG}
\end{figure}

The Directed Acyclic Graph (DAG) in Figure~\ref{figDAG} illustrates the structure of the Bayesian model used for estimation of the 4P FA model from Equations~\eqref{eq:3P-FAv1},~\eqref{eq:Zi},~\eqref{eq:4P-FA}. 

At the first level, the latent variable $\theta$ is assumed to follow a standard normal distribution: 
\begin{align}\label{theta_distr}
    \theta \sim \mathsf{N}(0, 1).
\end{align}
At the second level, latent binary responses $Z_i$ are modeled with Bernoulli distributions. Their success probabilities are modeled using the cumulative distribution function (CDF) corresponding to the distribution of the error terms in the FA model. The choice of the logistic CDF leads to the 4PL FA model: 

\begin{align}\label{Z_distrL}
    Z_i | \theta \sim Bernoulli\left(F_{logist} \left(\frac{\alpha_{i}\theta - \tau_{i}}{u_{i}}\right)\right) \; \; i = 1, \dots, m,
\end{align}
while using the normal CDF yields the 4PNO FA model:
\begin{align}\label{Z_distrN}
    Z_i | \theta \sim Bernoulli\left(F_{normal} \left(\frac{\alpha_{i}\theta - \tau_{i}}{u_{i}}\right)\right)\; \; i = 1, \dots, m.
\end{align}

At the third level, the observed binary responses $Y_i$ are also modeled using Bernoulli distributions. Their probabilities are determined by the state of the latent indicator $Z_i$: 

\begin{align}\label{Y_distr}
    Y_i | Z_i \sim Bernoulli\left(d_i^{Z_i} \cdot c_i^{1 - Z_i}\right) \; \; i = 1, \dots, m.
\end{align}

The model includes item-specific parameters $\alpha_i$, $\tau_i$, $c_i$, and $d_i$, for which we specify prior distributions as follows: 
\begin{align}\label{param_distr}
    \tau_i &\sim \mathsf{N}(0,1)\nonumber\\
    \alpha_i &\sim \mathsf{TruncNormal}(0.25,1, (0,1))\nonumber\\
    c_i &\sim \text{U}(0,1)\nonumber\\
    d_i &\sim \text{U}(c_i,1) \; \; i = 1, \dots, m.
\end{align}

These priors are chosen to reflect weakly informative assumptions, enabling flexible estimation while avoiding identifiability issues or unrealistic parameter values.

\vspace{1em}
%-----------------
\subsubsection{IRT parameterization}\label{typical_irt}
Typical parameterization is to directly model the probability of a correct response from Equation~\eqref{eq:4P-IRT}. In addition, difficulty and discrimination are used instead of factor analysis parameters. That approach is more straightforward, and the whole model is continuous in the Bayesian sense, which allows more advanced and faster samplers to be used. However, such parameterization avoids the discrete level and has a different interpretation not compatible with the factor analytic framework. One of the typical IRT parameterizations is the normal distribution for the latent variable:

\begin{align*}
    \theta \sim \mathsf{N}(0, 1),
\end{align*}

and Bernoulli distribution for observed answers:

\begin{align*}
    Y_i | \theta \sim Bernoulli\left(c_i + (d_i - c_i) \cdot F_{logist}\left[ a_i (\theta - b_i)  \right] \right) \; \; i = 1, \dots, m.
\end{align*}

There are various choices for difficulty and discrimination priors and very typical Beta priors for guessing and inattention; one choice inspired by \cite{culpepper2016revisiting} can be:

\begin{align*}
    a_i &\sim  \mathsf{TruncNormal}(2, 1/2, (0,\infty))\\
    b_i &\sim \mathsf{N}(0,1)\\
    c_i &\sim \text{Beta}(2,8)\\
    d_i &\sim \text{Beta}(8,2) \; \; i = 1, \dots, m.
\end{align*}

%-----------------
\subsection{Discrete latent responses}
The unobserved discrete latent variables $Z_i$ represent the true item-level performance of the respondents, adjusted for guessing and inattention. For clarity, we refer to their estimated values as NGNI (non-guessing and not-inattentive) item scores and denote them by $\hat{Z}_{i,p}$ where $p=1,\dots,n$ indexes persons and $i$ indexes items. The NGNI total score $\overline{Z}_p$ is then computed as the sum of $\hat{Z}_{i,p}$ across all items for each person $p$. We estimate these values using Bayesian methods, specifically as posterior expected values. While the true latent responses $Z_i$ are binary ($0$ or $1$), their estimates $\hat{Z}_{i,p}$ are continuous values in the interval $[0,1]$, representing the probability that the person truly solved the item. 

The NGNI total scores are expected to approximate the observed total scores, but the alignment varies, depending on the difficulty of the items solved. For example, solving four difficult items carries a different diagnostic value than solving four very easy ones. 

In addition to NGNI scores, we may estimate scores purified from only one of the two distortions. By fitting a 3P model (excluding inattention), we obtain NG (non-guessing) scores. Alternatively, using a model that includes only inattention yields NI (non-inattention) scores. 

Intuitively, NG scores tend to be lower than observed scores, since they account for overestimation due to guessing, while NI total scores tend to be higher, adjusting for underperformance due to inattention. We explore these properties using real data in Section~\ref{realdata}.

%-----------------
\subsection{Implementation in \texttt{R} and \texttt{jags}}\label{Rimplementation}

The estimation of the 4P FA model is implemented in R \parencite{Rko}, using the \texttt{runjags} package, which interfaces with \texttt{JAGS} \parencite{plummer2017jags}. 
This setup allows users to specify the hierarchical model described in Equations~\eqref{theta_distr}, \eqref{Z_distrL} or \eqref{Z_distrN}, \eqref{Y_distr}, and \eqref{param_distr}, as visualized in Figure~\ref{figDAG}. 

In addition to the model specification, initial values must be provided, which may slightly influence the resulting estimates. To address this, we performed two independent \texttt{JAGS} runs using different initial values. These runs can either be analyzed separately or combined to yield a single set of parameter estimates. In this work, we combine both chains to estimate the IRT parameters and NGNI scores. For each item-specific parameter ($\alpha$, $\tau$, $c$, $d$ or equivalently $a$, $b$, $c$, $d$), we report the posterior median as a point estimate. A complete implementation, including reproducible code, model scripts, and convergence diagnosis, is provided in the Electronic Supplementary Material.

%-----------------
\subsection{Implementation in \texttt{Python} and \texttt{pymc}}
The 4P FA model is also implemented in \texttt{Python} using the \texttt{pymc} package~\parencite{pymc2023}. While the modeling logic is similar to that of \texttt{JAGS}, the computational backend is fundamentally different. 
\texttt{pymc} relies on the No-U-Turn Sampler (NUTS)~\parencite[see][]{hoffman2014no}, an adaptive variant of Hamiltonian Monte Carlo (HMC), which offers greater efficiency for high-dimensional continuous models compared to traditional Gibbs sampling~\cite{casella1992explaining}, Metropolis-Hastings algorithms~\cite{chib1995understanding} or Hamilton sampler~\cite{betancourt2017conceptual}.

However, NUTS is only applicable for continuous variables. Since the 4P FA model includes discrete latent variables $Z_i$. \texttt{pymc} must use alternative samplers for this part of the model, which can substantially slow down computation. Despite these limitations, the \texttt{pymc} implementation provides a useful benchmark and flexibility for Python-based workflows. The code and details are available in the Electronic Supplementary Material. 

%----------------------------------------------------------------------------------------------

\section{Simulation study}\label{simstudy}
We performed a simulation study to demonstrate and verify the accuracy and stability of the proposed estimation method and to evaluate the practical implications of the FA parametrization.

\subsection{Simulation design}

\subsubsection{Data generation}
We chose the number of items $m$ to be $10$, $20$, $30$ and the number of respondents $n$ to be $50$, $100$, $2000$. Together, we have nine different scenarios; each was replicated $100$ times, so in total we had $900$ runs. The IRT parameters were simulated only once for each number of items as follows: 
The difficulty parameters $b$ were generated from $\texttt{N(0,1)}$, the discrimination parameters $a$ were chosen from the lognormal distribution $\texttt{logN(-0.3, 0.3)}$, the guessing $c$ and inattention $d$ parameters were generated from beta distributions, $\texttt{Beta(2,8)}$ and $\texttt{Beta(8,2)}$, respectively. Then they were scaled to ensure that $c<d$ was always satisfied. We intentionally did not use uniform distributions, a typical choice for modeling guessing and inattention, because we wanted to allow high guessing values, which are common in practical situations. Finally, we simulated the abilities $\theta$ from $\texttt{N(0,1)}$ during each replication. 

\subsubsection{Parameter estimation}
For each simulated dataset, the model parameters were estimated using the 4P FA model defined by equations~\eqref{eq:FA1}, \eqref{eq:Zi}, and~\eqref{eq:4P-FA}, with Bayesian samplers, as described in sections \ref{implementation}, with implementation in \texttt{R} as described in \ref{Rimplementation}. 
For comparison and reference, we used the typical IRT estimation of the model defined by equations~\eqref{eq:2P-IRT},~\eqref{eq:4P-IRT},~\eqref{2} using the EM algorithm, implemented in \texttt{R} using the \texttt{mirt} package. 
However, this could lead to comparing only the EM algorithm and Gibbs sampling, rather than the actual model. For that reason, we also added a third parameter-estimation method using JAGS in \texttt{R}, with the typical IRT parameterization described in Section~\ref{typical_irt}. 

As can be seen, the priors in the third estimation method are chosen very close to the actual simulation design and therefore are expected to perform best. 
To compare these three sets of estimates (for simplicity, we distinguish them as 4FA-B, 4IRT-EM, and 4IRT-B), we transformed all parameters into IRT-specific parameters: discriminations, difficulties, guessing, and inattention using Equations~\eqref{eq:eqvi}.

\subsubsection{Evaluation of the results}
For each run, we collected simulated IRT parameters and fitted parameters from each estimation method. %\texttt{mirt} the 4PL FA model and typical 4PL IRT parameterization. 
We investigated the convergence rates of %\texttt{mirt} 
the EM estimation algorithm, the convergence times of each method, and the mean-squared error calculated from all successful runs.

\subsection{Simulation results}
\paragraph{Convergence}
The IRT model estimates obtained using the EM algorithm did not converge in some cases or even diverged to absurdly high values. The default number of iterations was set to $2000$. When the EM algorithm failed to estimate the parameters before running out of iterations, or when the optimizer failed entirely, we considered it not converged. As the number of respondents increased, the convergence rate increased for smaller numbers of items ($n=10$ and $n=20$); see Table~\ref{tab_convergence}. For a larger number of respondents, we would need to increase the number of iterations.
%However, in the setups with $1000$ respondents, we observe the lowest convergence rates (see Table~\ref{tab_convergence}), which is in contradiction with intuition. We verified this effect on several runs and it was always present. Most probably because of the internal properties of \texttt{mirt} estimations. 
In the next part, we compare only runs that converged successfully with the EM algorithm. % \texttt{mirt}.

\begin{table}[!ht]
\centering
\begin{tabular}{rr|c}
\hline
n & m & Converged \\
\hline
50 & 10 & 22 \\
100 & 10 & 59 \\
2000 & 10 & 84 \\
\hline
50 & 20 & 17 \\
100 & 20 & 63 \\
2000 & 20 & 62 \\
\hline
50 & 30 & 26 \\
100 & 30 & 60 \\
2000 & 30 & 48 \\
\hline
\end{tabular}
    \caption{Comparison of successful convergence of 4IRT-EM %\texttt{mirt} 
    estimation for each scenario from the simulation study. Total number of estimations for each scenario is 100.}
    \label{tab_convergence}
\end{table}

\paragraph{Time efficiency}
For each run, we also record the time it took to estimate. For converged 4IRT-EM %\texttt{mirt} 
runs, the lengths are as expected. As the number of parameters increases, the required time increases. 
%Except for the setups with $1000$ respondents with \texttt{mirt} estimation due to the reasons described above. 
Although the average estimation time for 4IRT-EM is less than $5$ seconds, the average time of the JAGS estimates can reach thousands of seconds. For more details, see Table~\ref{tab_latency}

\begin{table}[htbp]
\centering

\label{tab:latency}
\begin{tabular}{rr|ccc}
\hline
n & m & 4FA-B & 4IRT-B & 4IRT-EM \\
\hline
50 & 10 & 11.725 & 10.855 & 0.740 \\
100 & 10 & 18.029 & 16.152 & 0.709 \\
2000 & 10 & 673.216 & 472.018 & 0.895 \\
\hline
50 & 20 & 19.816 & 18.055 & 1.494 \\
100 & 20 & 34.530 & 30.154 & 1.241 \\
2000 & 20 & 1370.070 & 969.755 & 1.771 \\
\hline
50 & 30 & 26.871 & 23.942 & 1.620 \\
100 & 30 & 49.839 & 41.059 & 2.204 \\
2000 & 30 & 2080.250 & 1475.428 & 4.571 \\
\hline
\end{tabular}
\caption{Comparison of average estimation times for each setup with n respondents and m items for each method in seconds. 4FA-B stands for the presented four-parameter factor analysis model estimated by the Bayesian framework, 4IRT is the four-parameter IRT model with typical IRT parameterization also estimated by the Bayesian framework, and 4IRT-EM is the four-parameter IRT model estimated by the EM algorithm.}
 \label{tab_latency}
\end{table}

% \begin{table}[!ht]
%     \centering
%     \includegraphics[scale=0.70]{figures/simulation/latency_table.pdf}
%      \vspace{-4mm}
%     \caption{Comparison of average time of estimation times for each setup with n respondents and m items for each method in seconds. 4FA-B stands for presented four parameter factor analysis model estimated by Bayesian framework, 4IRT is four parametric IRT model with typical IRT parameterization estimated by Bayesian framework, and 4IRT-EM is four parametric IRT model estimated by EM algorithm.}
%     \label{tab_latency}
% \end{table}

\begin{figure}[!ht]
    \centering
    \includegraphics[scale=0.62]{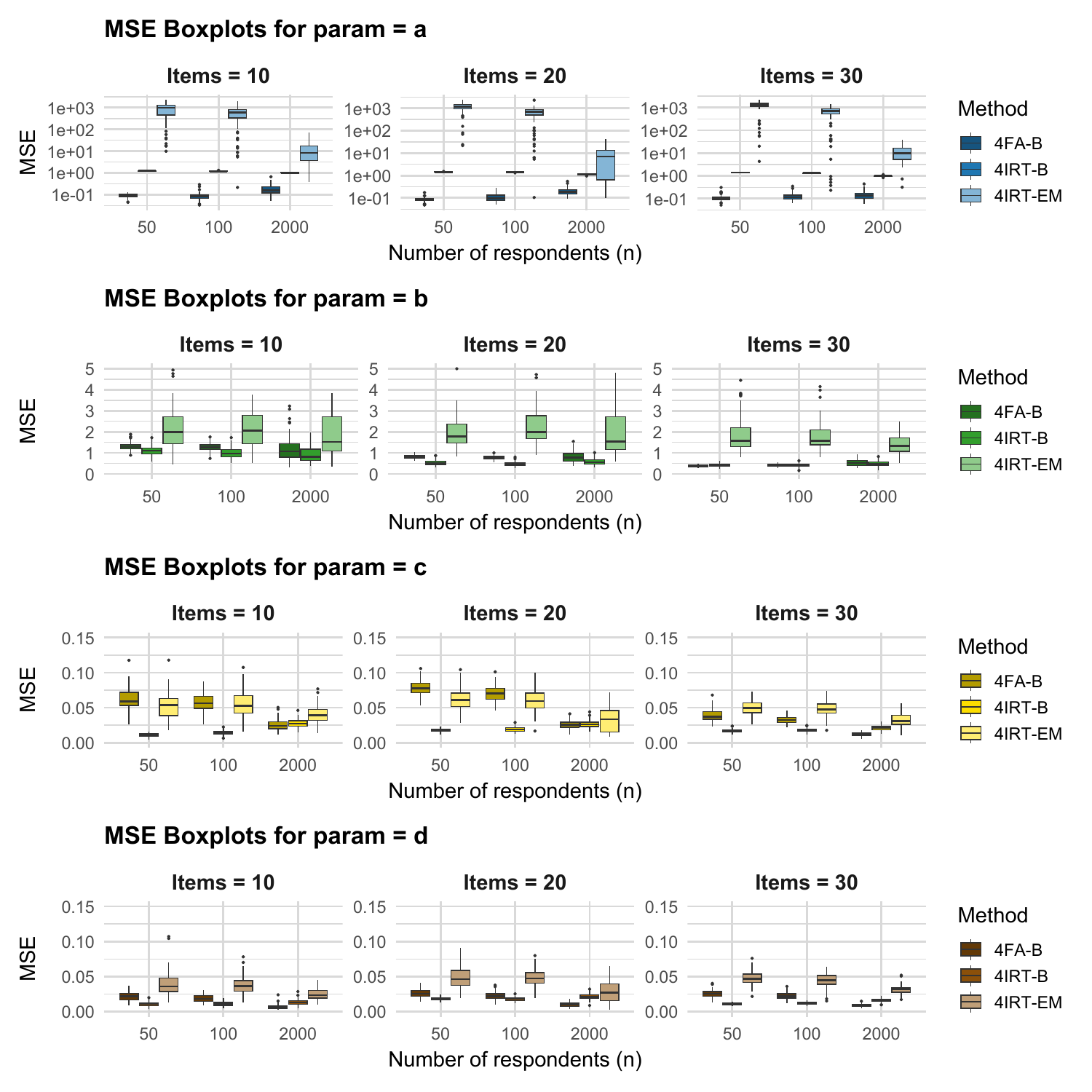}
     \vspace{-4mm}
    \caption{Boxplots of Mean Square Errors (MSE) for each scenario and parameter discrimination (a), difficulty (b), guessing (c), and inattention (d) from simulation study. From each parameter group of boxplots, the left is for the proposed 4PL FA model using Bayesian estimates from JAGS, the middle is for the typical 4PL IRT Bayesian parameterization using JAGS, and the right is for the IRT model fitted by the EM algorithm using the \texttt{mirt} package. Estimated values are compared to true simulated data. 4FA-B stands for the presented four-parameter factor analysis model estimated by the Bayesian framework, 4IRT is the four-parameter IRT model with typical IRT parameterization estimated by the Bayesian framework, and 4IRT-EM is the four-parameter IRT model estimated by the EM algorithm.}
    \label{box_sim_mean}
\end{figure}

\paragraph{Precision}
Although we filter only on converged runs, some MSE of 4IRT-EM estimates are outside of the reasonable range; see Boxplots~\ref{box_sim_mean}. For more numerical estimates of means and medians of MSE, see Tables~\ref{tab_sim_mean} and~\ref{tab_sim_median} in the Appendix.

However, another problem with the EM algorithm is the need for a sufficient sample size, especially in situations with only $50$ respondents. As seen in scenarios with an increasing number of respondents, precision increases and, therefore, the MSE decreases.
The smaller sample sizes were chosen intentionally to also demonstrate the advantages of Bayesian estimates in such situations. %Because they have prior knowledge that the parameters are reasonable.
Nevertheless, Bayesian estimates from the 4FA or 4IRT model outperform the frequentist EM approach even with larger sample sizes. This, of course, comes at a cost in computational time. For sample sizes such as $50$ or $100$ respondents, the model estimation took less than one minute. With a larger sample size, such as the $2000$ respondents, the estimation time is on the magnitude order of tens of minutes, due to the increasing number of parameters and latent variables. Accordingly, the number of samplings in the Gibbs sampler, used to fit the Bayesian model, needs to be increased in order to compensate for the increase in parameters. Otherwise, we cannot guarantee that 
the MSE will decrease as the sample size increases. This effect is evident in the estimates of the discrimination and difficulty parameters.

%----------------------------------------------------------------------------------------------

\section{Real data examples}\label{realdata}
 We demonstrate the estimation of the item parameters and NGNI scores under the proposed 4P FA model from Section~\ref{model_spec}, and we compare these with parameter estimates from the 4P IRT model~\eqref{eq:4P-IRT} estimated by both the EM algorithm and the Bayesian framework. 
To facilitate comparison, we convert FA parameters, i.e., loading, threshold, guessing, and inattention ($\alpha$, $\tau$, $c$, $d$), to their IRT equivalents, i.e., discrimination, difficulty, guessing, and inattention ($a$, $b$, $c$, $d$), using the relationships in Equation~\eqref{eq:eqvi}. Reverse transformations are also possible.

We focus on the logistic variants of both models (i.e., 4PL IRT and 4PL FA), where the IRC is defined by the logistic function and the FA model's error terms,  $\epsilon_{i}$, follow a logistic distribution. The same methodology applies to the normal-ogive version of the models. 

In addition to item parameters, we examine NGNI scores estimated from both \texttt{JAGS} and \texttt{pymc} implementations and compute NG and NI total scores, which isolate the effects of guessing and inattention, respectively.

\subsection{MSATB data}
\begin{figure}[!ht]
    \centering
    \includegraphics[scale=0.65]{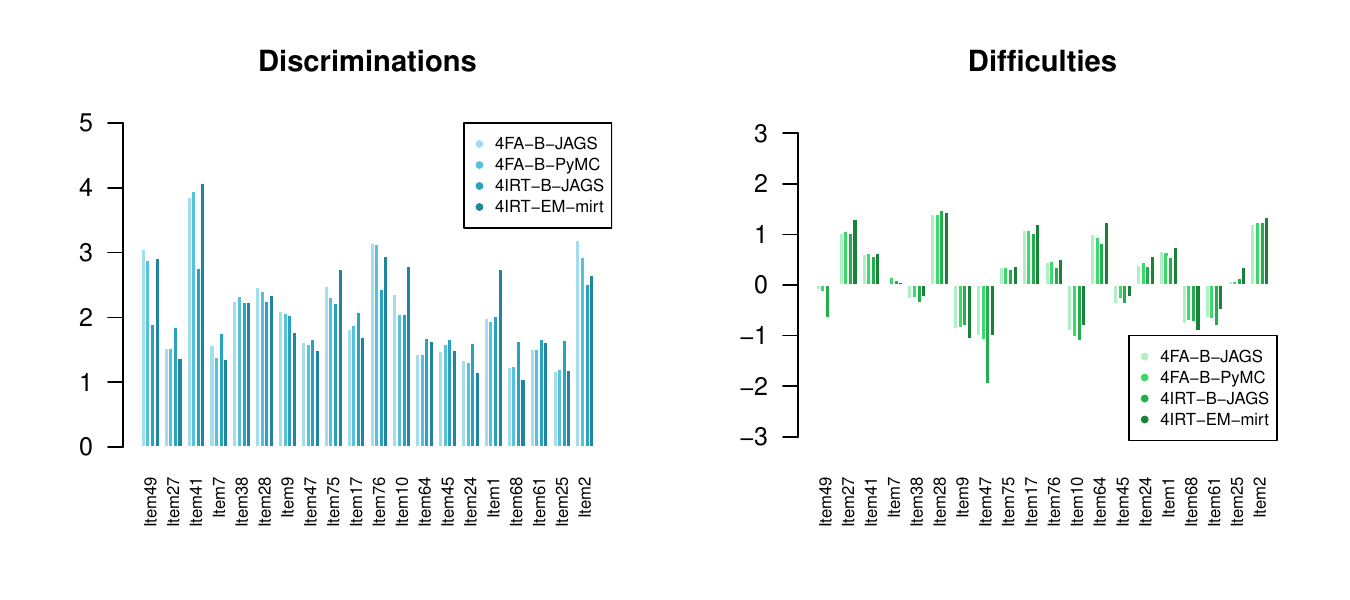}
    \includegraphics[scale=0.65]{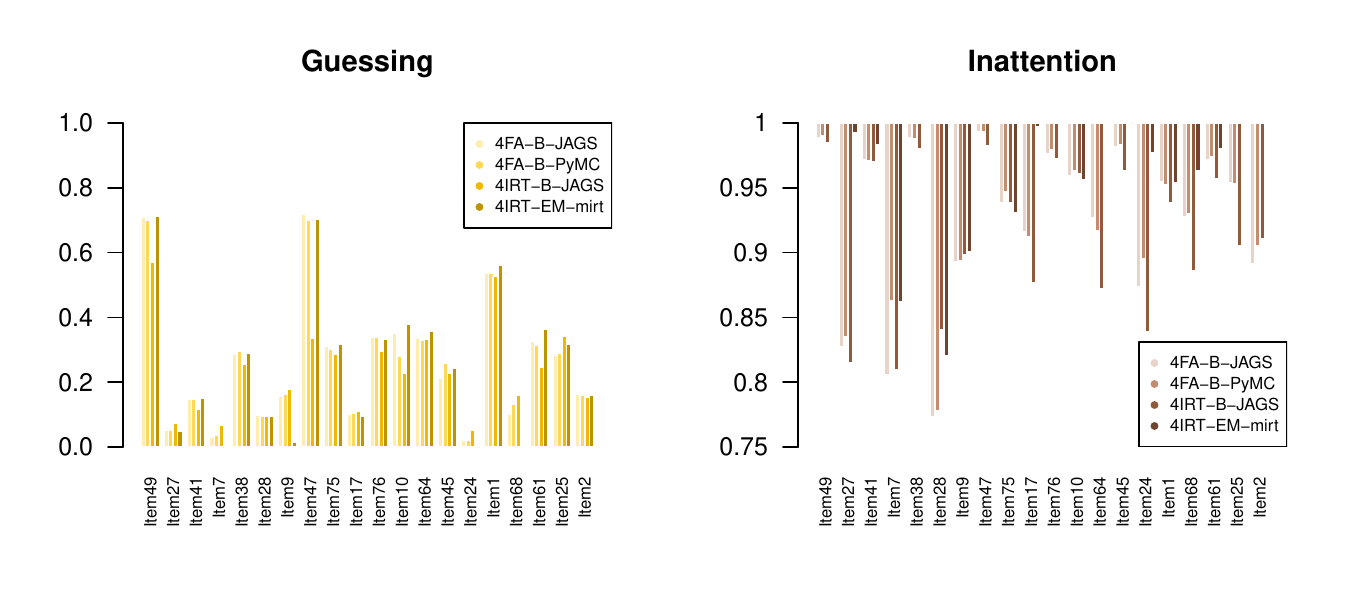}
     \vspace{-10mm}
    \caption{Comparison of estimated parameters (discrimination, difficulty, guessing, and inattention) from 4PL FA and 4PL IRT  for the MSATB dataset.}
    \label{fig1}
\end{figure}

The dataset contains 1407 responses to 20 dichotomous items. We compare parameter estimates from the 4PL FA model using \texttt{JAGS} in \texttt{R} and \texttt{pymc} in \texttt{Python} against estimates from the 4PL IRT model via the \texttt{mirt} package \parencite{chalmers2012mirt} and via \texttt{JAGS} using typical IRT parameterization from Section~\ref{typical_irt}.
The estimated parameters from both Bayesian and EM frameworks show strong agreement. Notably, inattention parameters under the FA model tend to be slightly lower (further from~1) than the IRT estimates, possibly due to the uniform prior on inattention, which allows greater variability. Bayesian estimations of discrimination parameters using FA parameterization are more aligned with IRT estimates than Bayesian estimates using IRT parameterization. A similar effect can be observed in the estimation of guessing parameters.
 See Figure~\ref{fig1}, Appendix Table~\ref{tab1}, and MSE comparison Table~\ref{mse_mirt_tab_1}.
 
Like the 4PL IRT, the 4PL FA model also provides estimates of latent abilities. However, it also enables the computation of NGNI scores, which adjust for guessing and inattention--capabilities not available in standard IRT frameworks. Table~\ref{tabNGNI1} shows the NGNI scores for the first six respondents.

To further explore these effects, we also compute NG and NI scores individually. Implementation details are provided in the Supplementary Material.
\begin{table}[h!]
\centering
\begin{tabular}{r|rrrrrr}
\hline
Respondent & Total & NGNI total & NGNI total & NG total & NI total \\
 & score & score pymc & score JAGS & score JAGS & score JAGS \\
\hline
1 & 9 & \textbf{6.5050} & \textbf{6.6045} & \textbf{6.5968} & \textbf{9.3583} \\
2 & 14 & 14.0880 & 14.2138 & 12.2822 & 14.7718 \\
3 & 9 & \textbf{5.2670} & \textbf{5.1365} & \textbf{5.5493} & \textbf{9.3637} \\
4 & 11 & 9.2210 & 9.2475 & 8.9025 & 11.4463 \\
5 & 17 & 16.9020 & 17.0858 & 15.9277 & 17.4650 \\
6 & 13 & 11.6500 & 11.6938 & 10.9482 & 13.5310 \\
\hline
\end{tabular}
\caption{Comparison of observed total scores, non-guessing and not-inattention (NGNI) scores, non-guessing (NG) scores, and non-inattention (NI) scores for the first six respondents (MSATB dataset). NGNI scores account for item severity and remove the effects of guessing and inattention.}
\label{tabNGNI1}
\end{table}
As in standard IRT models, respondents with the same total score can have different NGNI scores depending on their response patterns. Moreover, even respondents with identical response patterns may show slight variations in NGNI scores due to stochasticity in Bayesian estimation. Table~\ref{tabNGNI2} illustrates this effect for respondents who correctly answered all items. This variation can be removed by averaging NGNI total scores across respondents with the same response pattern; see Table~\ref{tabNGNI3} in Appendix~\ref{sec:tabs}.
\begin{table}[h!]
\centering
\begin{tabular}{l|rrrrrrr}
Respondent & 498 & 895 & 1022 & 1093 & 1102 & 1160 & 1220\\
\hline
NGNI & 19.6398 & 19.6368 & 19.6342 & 19.6653 & 19.6467 & 19.6577 & 19.6542 \\
\end{tabular}
\caption{NGNI total scores for respondents with perfect observed scores (20/20) in the MSATB dataset. Small differences arise due to Bayesian sampling variability.}
\label{tabNGNI2}
\end{table}

\subsection{Anxiety data}
This dataset contains responses from 766 participants on 29 Likert-type items. We dichotomize the responses such that "Never" = 0 and all other options = 1. As with MSATB, we compare the 4PL FA estimates from \texttt{JAGS} and \texttt{pymc} to 4PL IRT estimates from \texttt{mirt} using EM algorithm and typical IRT parameterization from Section~\ref{typical_irt} using JAGS.

\begin{figure}[!h]
    \centering
    \includegraphics[scale=0.60]{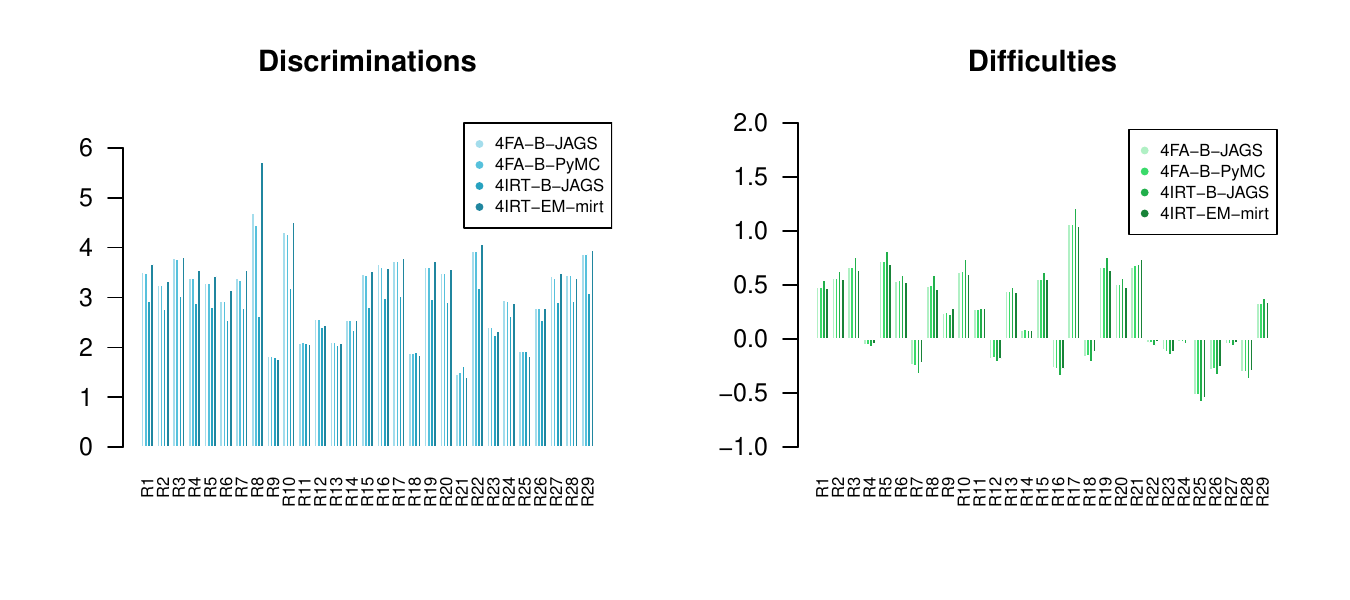}
    \includegraphics[scale=0.60]{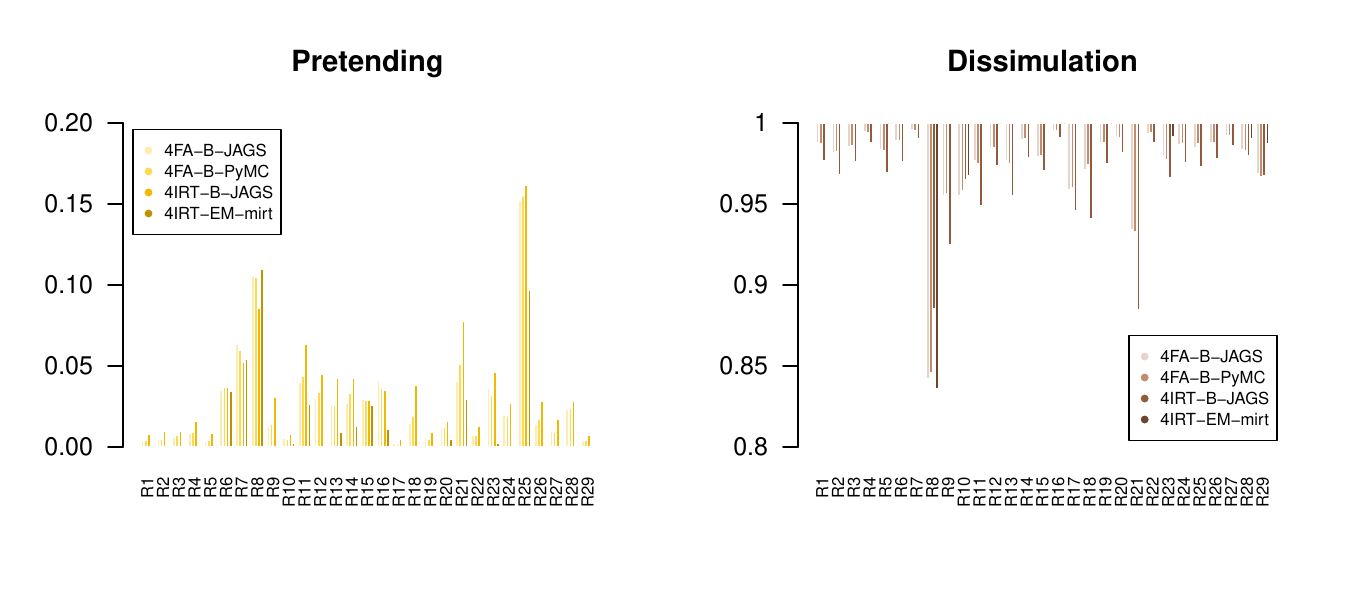}
     \vspace{-10mm}
    \caption{Comparison of estimated item parameters (discrimination, difficulty, pretending, and dissimulation) from 4PL FA model and 4PL IRT models for the Anxiety dataset.
    }
    \label{fig3}
\end{figure}

Estimates of typical IRT parameterization 4IRT-B-JAGS tend to differ from other estimation methods for all parameters. Otherwise, estimated parameters are closely aligned across frameworks, with minor exceptions. For example, the discrimination parameter for item $R8$ is higher in the IRT model ($5.72$) than in FA-based estimates ($4.69$ for JAGS, $4.44$ for pymc). The pretending parameters are consistent across all models, whereas the dissimulation parameters in the FA model again tend to diverge more from 1 compared to IRT estimates (e.g., items R17 and R21). See Figures~\ref{fig3}, Appendix Table~\ref{tab2}, and Table~\ref{mse_mirt_tab_2} for detailed comparisons. 

We also compute NGNI scores for each respondent. These scores remain on the same scale as observed total scores but are corrected for guessing and inattention. Table~\ref{tabNGNI_anx_1} shows the results for the first six respondents.

\begin{table}[h!]
\centering
\begin{tabular}{r|rrrr}
\hline
Respondent & Total score & NGNI total score JAGS & NGNI total score pymc\\
\hline
1 & 11 & \textbf{9.8067} & \textbf{9.8210} \\
2 & 1 & 0.5778 & 0.5690 \\
3 & 11 & \textbf{10.7903} & \textbf{10.7520} \\
4 & 7 & 6.8105 & 6.7880 \\
5 & 0 & 0.0183 & 0.0170 \\
6 & 10 & 9.9972 & 10.0100 \\
\hline
\end{tabular}
\caption{Observed and NGNI scores for the first six respondents in the Anxiety dataset. NGNI scores are adjusted for guessing and inattention.}
\label{tabNGNI_anx_1}
\end{table}

%----------------------------------------------------------------------------------------------
\section{Discussion}\label{discussion}

In this article, we proposed a four-parameter factor-analytic (4P FA) model and demonstrated both analytically and empirically its equivalence to the 4P IRT model. Our work extends the well-established literature on the relationship between FA and IRT, especially building on the equivalence shown for dichotomized 2P models \parencite{takane1987relationship}. While prior research has explored more complex extensions with multiple latent variables or covariates (see~\cite{moustaki2004factor}), the 3P and 4P cases have received limited attention in the FA framework. 
By introducing a second-level mixture component, we established the equivalence between the 4PNO FA and 4PNO IRT models (and their logistic analogs 4PL, as well as restricted 3P versions). This hierarchical mixture structure builds on the dichotomization step in standard FA-IRT equivalencies and yields a clear interpretation of model components. More broadly, our results extend the classical FA--IRT correspondence beyond the two-parameter case and demonstrate that estimates from one framework can be natively transformed to parameters of the other framework. 

We formally proved the equivalence of the proposed 4P FA model with the 4P IRT model under specific assumptions. 
We also developed a Bayesian estimation approach and implemented it in both \texttt{R} and \texttt{Python}, showing that the estimates align closely with those obtained from the \texttt{mirt} package in \texttt{R}. We performed a simulation study in which the proposed Bayesian estimate under the 4PL FA model specification outperformed the EM estimates and, in some cases, even the Bayesian estimate under the 4PL IRT model specification. Superior performance of the Bayesian estimates was expected for smaller sample sizes, but less so for large ones, where we expected the IRT model with the EM-algorithm to be more precise. Given that the data were simulated from the 4PL IRT model with IRT parameterization, the Bayes estimates under the IRT model specification, and with priors close to those of the simulated data, can be considered a benchmark of estimation methods. However, the Bayes estimates under the 4PL FA model specification were sometimes more precise than those under the 4PL IRT model, especially with larger numbers of items and respondents, supporting the 4PL FA modeling alternative. %, see Boxplots~\ref{box_sim_mean}. 
One possible explanation is that the prior distribution for loading is truncated to the bounded interval $(0,1)$, while the typical prior distribution for discrimination requires an unbounded distribution.
% the FA parameterization induces a different posterior geometry and dependence structure among parameters, which may improve sampling efficiency or provide a degree of regularization. 
The mechanism underlying this advantage deserves further investigation and may offer insights into the estimation of higher-parameter latent variable models more generally. 
While traditional FA estimation methods typically rely on a variance-covariance matrix, assuming underlying normal or logistic distributions, the 4P FA model requires different assumptions. Extending such classical approaches to this setting remains an important direction for future research. 

Although our primary contribution is situated within the FA framework, our estimation approach complements existing frequentist and Bayesian methods for 4P IRT models. For example, \cite{hladka2026newIterativeAlgorithms} introduced an iterative algorithm for the hierarchical 4PL model involving hypothetical response statuses. 
Other researchers, such as~\cite[]{beguin2001mcmc} and ~\cite[]{sahu2002bayesian}, developed Bayesian approaches for the 3PNO IRT model using binary latent variables. 
\cite{fu2021gibbs} introduced a similar augmentation strategy, but 
with auxiliary Bernoulli and uniform variables. \cite{culpepper2016revisiting, culpepper2017prevalence} proposed a Gibbs sampling framework for 4PNO IRT models, where discrete augmented variables resemble our latent responses $Z_i$. 
\cite{zheng2021fast, meng2020marginalized, meng2023mixed} employed EM-type algorithms with different parameter structures. 
Besides primarily assuming the IRT framework, these prior studies differ from our work in either the hierarchical structure or/and analytical focus. For example, \cite{hladka2026newIterativeAlgorithms} and \cite{fu2021gibbs} employ latent variables at different levels of the modeling hierarchy, while others, such as \cite{beguin2001mcmc} and \cite{culpepper2017prevalence}, focus more on deriving posterior distributions or comparing 4PNO and 3PNO IRT models. In contrast, our work is grounded in the FA framework and emphasizes the formal equivalence between 4P FA and 4P IRT models, supported by mathematical proofs and empirical validation.

Beyond item parameters, the 4P FA model yields the NGNI item scores---posterior expectations of the discrete latent responses $Z_i$---and their sums, the NGNI total scores $\overline{Z}_p$. Similar to individual latent ability estimates $\hat{\theta}_p$, NGNI scores account for item difficulty and discrimination while removing the influence of guessing and inattention. However, unlike $\hat{\theta}_p$, NGNI total scores are bounded within an interpretable scale, ranging from 0 to the total number of items, making them easier to communicate and more directly comparable to observed scores. This property may be particularly attractive in applied settings, where reporting adjusted scores on the original score metric is often preferable to reporting latent trait estimates on an arbitrary standardized scale. 

%We advocate for using posterior means for NGNI scores, rather than medians, due to their continuous nature. Posterior medians are more robust for estimating item parameters, but in the case of binary latent responses, medians collapse to either 0 or 1 and thus lose interpretive power. Alternatively, NGNI scores can be approximated by inserting point estimates of $\theta$ into Equations~\eqref{Z_distrL} or~\eqref{Z_distrN}, but this does not yield true posterior expectations due to the nonlinearity of the CDFs (cf. Jensen's inequality; \cite{cohn2013measure}).

One limitation of Bayesian estimation is that respondents with identical response patterns may receive slightly different estimates due to sampling variability. This is also true for latent abilities $\theta$, although deterministic alternatives exist for them. To mitigate this, we propose a heuristic to average the NGNI scores across identical response patterns.

Another consideration is the specification of prior distributions. While we used relatively uninformative priors (uniform and truncated normal), alternative parameterizations exist. For instance, in the IRT framework,~\cite{burkner2020analysing} 
used transformations (e.g., $exp()$, $logit()$) with normal priors. 
Other works, such as ~\cite{loken2010estimation, waller2010measuring} employed beta priors for guessing and inattention.  Our choice of truncated normal priors for factor loadings and standard normal priors for thresholds was motivated by performance and comparability to EM estimates. 

Several avenues remain for further work. First, more informative or empirically driven priors may improve estimation quality. Second, extensions to include covariates could allow for modeling group-level differences in guessing or inattention. While multigroup FA models raise issues of identifiability, they present a promising direction for enhancing measurement invariance analysis. 

Despite its limitations, our work makes several important contributions. We introduce and validate the 4P FA model as a practically useful extension of FA, bridging it with 4P IRT models; we also offer a Bayesian estimation method and demonstrate improved estimation stability and parameter recovery in a simulation study, particularly in settings where the EM algorithm fails to converge. The NGNI scores, constrained to the same scale as total scores but cleaned of guessing and inattention, offer a valuable interpretive tool. More broadly, the proposed model strengthens the theoretical bridge between two major traditions of latent variable modeling and illustrates how concepts traditionally associated with IRT can be incorporated into a factor-analytic framework. We hope these results foster a closer integration of the IRT and FA approaches, even in more advanced models.

%--------------------------------------------------
% ACKNOWLEDGEMENT
%--------------------------------------------------
\section*{Acknowledgement}

%[Blinded for review.]
The work was funded by the Czech Science Foundation grant number 25-16951S, project ``Research of Excellence on Digital Technologies and Wellbeing CZ.02.01.01/00/22\_008/0004583" which is co-financed by the European Union, and by the institutional support RVO 67985807.  %We are thankful to anonymous reviewers for helpful comments and suggestions on earlier versions of the manuscript. 
In preparing this manuscript, ChatGPT-4 (OpenAI, 2026) was used for 
language refinement. All content, including intellectual contributions, analysis, and interpretation, originates from the authors. All AI-generated suggestions were reviewed, edited, and approved by the authors, who take full responsibility for the final content.

%--------------------------------------------------
% SUPPLEMENTARY MATERIAL
%--------------------------------------------------
\section*{Online Supplementary Material}

Electronic Supplementary Material, including the accompanying \texttt{R} and \texttt{Python} scripts, is available at \href{https://osf.io/mr7pz/view_only=ff4515f5ca8b411f86b3d337d5dba309}{https://osf.io/mr7pz/?view\_only=ff4515f5ca8b411f86b3d337d5dba309}.

\printbibliography

@book{bartholomew2008analysis,
  title={Analysis of Multivariate Social Science Data},
  author={Bartholomew, David J and Steele, Fiona and Moustaki, Irini},
  year={2008},
  publisher={CRC Press}
}

@article{barton1981upper,
  title={An upper asymptote for the three-parameter logistic item-response model},
  author={Barton, Mark A and Lord, Frederic M},
  journal={ETS Research Report Series},
  volume={1981},
  number={1},
  pages={1--8},
  year={1981},
  publisher={Wiley Online Library},
  doi={10.1002/j.2333-8504.1981.tb01255.x}
}

@article{battauz2020regularized,
    author = {Battauz, Michela},
    title = {Regularized Estimation of the Four-Parameter Logistic Model},
    journal = {Psych},
    volume = {2},
    year = {2020},
    number = {4},
    pages = {269--278},
    doi = {10.3390/psych2040020}
}

@article{beguin2001mcmc,
  title={{MCMC} estimation and some model-fit analysis of multidimensional {IRT} models},
  author={B{\'e}guin, Anton A and Glas, Ceec AW},
  journal={Psychometrika},
  volume={66},
  pages={541--561},
  year={2001},
  publisher={Springer},
  doi={10.1007/BF02296195}
}

@article{betancourt2017conceptual,
  title={A conceptual introduction to Hamiltonian \\ Monte Carlo},
  author={Betancourt, Michael},
  journal={arXiv preprint arXiv:1701.02434},
  year={2017},
  doi={10.48550/arXiv.1701.02434}
}

@incollection{birnbaum1968statistical,
  booktitle={Statistical theories of mental test scores},
  author={Birnbaum, Alfred},
  editor={Lord, Frederic M and Novick, Melvin R},
  title={Some latent trait models and their use in inferring an examinee's ability},
  publisher={Addison-Wesley, Reading, MA},
  year={1968},
  pages={397--479}
}

@article{burkner2020analysing,
  title={Analysing standard progressive matrices ({SPM-LS}) with {Bayesian} item response models},
  author={B{\"u}rkner, Paul-Christian},
  journal={Journal of Intelligence},
  volume={8},
  number={1},
  pages={5},
  year={2020},
  publisher={MDPI},
  doi={10.3390/jintelligence8010005}
}

@article{casella1992explaining,
  title={Explaining the {Gibbs} Sampler},
  author={Casella, George and George, Edward I},
  journal={American Statistician},
  pages={167--174},
  year={1992},
  publisher={JSTOR},
  doi={10.2307/2685208}
}

@article{chalmers2012mirt,
  title={{mirt:} A multidimensional item response theory package for the {R} environment},
  author={Chalmers, R Philip},
  journal={Journal of Statistical Software},
  volume={48},
  pages={1--29},
  year={2012},
  doi={10.18637/jss.v048.i06}
}

@article{chib1995understanding,
  title={Understanding the {Metropolis}-{Hastings} algorithm},
  author={Chib, Siddhartha and Greenberg, Edward},
  journal={The American Statistician},
  volume={49},
  number={4},
  pages={327--335},
  year={1995},
  publisher={Taylor \& Francis},
  doi={10.1080/00031305.1995.10476177}
}

@book{cohn2013measure,
  title={Measure Theory},
  author={Cohn, Donald L},
  volume={5},
  year={2013},
  publisher={Springer}
}

@article{culpepper2016revisiting,
  title={Revisiting the 4-parameter item response model: Bayesian estimation and application},
  author={Culpepper, Steven Andrew},
  journal={Psychometrika},
  volume={81},
  number={4},
  pages={1142--1163},
  year={2016},
  publisher={Springer},
  doi={10.1007/s11336-015-9477-6}
}

@article{culpepper2017prevalence,
  title={The prevalence and implications of slipping on low-stakes, large-scale assessments},
  author={Culpepper, Steven Andrew},
  journal={Journal of Educational and Behavioral Statistics},
  volume={42},
  number={6},
  pages={706--725},
  year={2017},
  publisher={SAGE Publications Sage CA: Los Angeles, CA},
  doi={10.3102/1076998617705653}
}

@article{fu2021gibbs,
  title={A {Gibbs} sampler for the multidimensional four-parameter logistic item response model via a data augmentation scheme},
  author={Fu, Zhihui and Zhang, Susu and Su, Ya-Hui and Shi, Ningzhong and Tao, Jian},
  journal={British Journal of Mathematical and Statistical Psychology},
  volume={74},
  number={3},
  pages={427--464},
  year={2021},
  publisher={Wiley Online Library},
  doi={10.1111/bmsp.12234}
}

@article{haley1952estimation,
  title={Estimation of the dosage mortality when the dose is subject to error (Technical Report, No. 15)},
  author={Haley, DC},
  year={1952},
  publisher={Stanford, CA: Stanford University, Applied Mathematics and Statistics Laboratory}
}

@article{hladka2026newIterativeAlgorithms,
  title={New Iterative Algorithms for Estimation of Item Functioning},
  author={Hladk{\'a}, Ad{\'e}la and Martinkov{\'a}, Patr{\'\i}cia and Brabec, Marek},
  journal={Journal of Educational and Behavioral Statistics},
  year={2026},
  volume={51},
  number={1},
  pages={175--205},
  doi={10.3102/10769986241312354}
}

@article{hoffman2014no,
  title={The {No-U-Turn} Sampler: Adaptively setting path lengths in {Hamiltonian Monte Carlo}},
  author={Hoffman, Matthew D and Gelman, Andrew and others},
  journal={Journal of Machine Learning Research},
  volume={15},
  number={1},
  pages={1593--1623},
  year={2014}
}

@article{kamata2008note,
  title={A note on the relation between factor analytic and item response theory models},
  author={Kamata, Akihito and Bauer, Daniel J},
  journal={Structural Equation Modeling: A Multidisciplinary Journal},
  volume={15},
  number={1},
  pages={136--153},
  year={2008},
  publisher={Taylor \& Francis},
  doi={10.1080/10705510701758406}
}

@article{loken2010estimation,
  title={Estimation of a four-parameter item response theory model},
  author={Loken, Eric and Rulison, Kelly L},
  journal={British Journal of Mathematical and Statistical Psychology},
  volume={63},
  number={3},
  pages={509--525},
  year={2010},
  publisher={Wiley Online Library},
  doi={10.1348/000711009X474502}
}

@incollection{lord2008statistical,
  booktitle={Statistical Theories of Mental Test Scores},
  author={Lord, Frederic M and Novick, Melvin R},
  year={1968},
  pages={358--394},
  year={1967},
  publisher={IAP}
}

@book{martinkova2023computational,
  title={Computational Aspects of Psychometric Methods: With {R}},
  author={Martinková, Patrícia and Hladká, Adéla},
  year={2023},
  publisher={CRC Press},
  doi={10.1201/9781003054313}
}

@book{mendenhall2012introduction,
  title={Introduction to probability and statistics},
  author={Mendenhall, William and Beaver, Robert J and Beaver, Barbara M},
  pages={153},
  year={2012},
  publisher={Cengage Learning}
}

@article{meng2020marginalized,
    title = {Marginalized maximum a posteriori estimation for the four-parameter logistic model under a mixture modelling framework},
    author = {Meng, Xiangbin and Xu, Gongjun and Zhang, Jiwei and Tao, Jian},
    journal = {British Journal of Mathematical and Statistical Psychology},
    volume = {73},
    pages = {51--82},
    year = {2020},
    publisher = {Wiley Online Library},
    doi = {10.1111/bmsp.12185}
}

@article{meng2023mixed,
  title={A mixed stochastic approximation EM (MSAEM) algorithm for the estimation of the four-parameter normal ogive model},
  author={Meng, Xiangbin and Xu, Gongjun},
  journal={psychometrika},
  volume={88},
  number={4},
  pages={1407--1442},
  year={2023},
  publisher={Cambridge University Press \& Assessment}
}

@book{McDonald1999,
year = {1999},
author = {McDonald, Roderick P.},
address = {Mahwah, N.J},
booktitle = {Test theory: A unified treatment},
isbn = {0-8058-3075-8},
language = {eng},
publisher = {Lawrence Erlbaum Associates},
title = {Test theory : a unified treatment },
}

@article{moustaki2004factor,
  title={Factor models for ordinal variables with covariate effects on the manifest and latent variables: A comparison of LISREL and IRT approaches},
  author={Moustaki, Irini and J{\"o}reskog, Karl G and Mavridis, Dimitris},
  journal={Structural Equation Modeling},
  volume={11},
  number={4},
  pages={487--513},
  year={2004},
  publisher={Taylor \& Francis},
  doi={10.1207/s15328007sem1104_1}
}

@article{noventa2019generalization,
  title={On a generalization of local independence in item response theory based on knowledge space theory},
  author={Noventa, Stefano and Spoto, Andrea and Heller, J{\"u}rgen and Kelava, Augustin},
  journal={Psychometrika},
  volume={84},
  pages={395--421},
  year={2019},
  publisher={Springer},
  doi={10.1007/s11336-018-9645-6}
}

@article{plummer2017jags,
  title={JAGS Version 4.3. 0 user manual},
  author={Plummer, Martyn},
  year={2017}
}

@book{poirier1995intermediate,
  title={Intermediate statistics and econometrics: A comparative approach},
  author={Poirier, Dale J},
  year={1995},
  publisher={MIT Press}
}

@article{pymc2023,
  title={PyMC: a modern, and comprehensive probabilistic programming framework in {Python}},
  author={Abril-Pla, Oriol and Andreani, Virgile and Carroll, Colin and Dong, Larry and Fonnesbeck, Christopher J and Kochurov, Maxim and Kumar, Ravin and Lao, Junpeng and Luhmann, Christian C and Martin, Osvaldo A and others},
  journal={PeerJ Computer Science},
  volume={9},
  pages={e1516},
  year={2023},
  publisher={PeerJ Inc.},
  doi={10.7717/peerj-cs.1516}
}

@book{python,
 author = {Van Rossum, Guido and Drake, Fred L.},
 title = {Python 3 Reference Manual},
 year = {2009},
 isbn = {1441412697},
 publisher = {CreateSpace},
 address = {Scotts Valley, CA}
}

@book{rao2007psychometrics,
  title={Psychometrics},
  author={Rao, Calyampudi Radhakrishna and Sinharay, Sandip},
  series={Handbook of Statistics},
  volume={26},
  year={2007},
  edition={First},
  publisher={Elsevier},
  address = {Amsterdam; Boston}
}

@manual{Rko,
  title = {R: A Language and Environment for Statistical Computing},
  author = {{R Core Team}},
  note = {Computer software manual},
  organization = {R Foundation for Statistical Computing},
  address = {Vienna, Austria},
  year = {2025},
  url = {https://www.R-project.org/},
}

@article{sahu2002bayesian,
  title={Bayesian estimation and model choice in item response models},
  author={Sahu, Sujit K},
  journal={Journal of Statistical Computation and Simulation},
  volume={72},
  number={3},
  pages={217--232},
  year={2002},
  publisher={Taylor \& Francis},
  doi={10.1080/00949650212387}
}

@article{takane1987relationship,
  title={On the relationship between item response theory and factor analysis of discretized variables},
  author={Takane, Yoshio and De Leeuw, Jan},
  journal={Psychometrika},
  volume={52},
  number={3},
  pages={393--408},
  year={1987},
  publisher={Springer},
  doi={10.1007/BF02294363}
}

@incollection{waller2010measuring,
  author      = "Waller, Niels G and Reise, Steven P",
  title       = "Measuring psychopathology with nonstandard item response theory models: Fitting the four-parameter model to the Minnesota Multiphasic Personality Inventory.",
  booktitle   = "Measuring psychological constructs: Advances in model-based approaches",
  publisher   = "American Psychological Association",
  year        = 2010,
  pages       = "147-173",
  chapter     = 7,
}

@article{waller2017bayesian,
  title={Bayesian modal estimation of the four-parameter item response model in real, realistic, and idealized data sets},
  author={Waller, Niels G and Feuerstahler, Leah},
  journal={Multivariate Behavioral Research},
  volume={52},
  number={3},
  pages={350--370},
  year={2017},
  publisher={Taylor \& Francis},
  doi={10.1080/00273171.2017.1292893}
}

@article{zheng2021fast,
  title={Fast {Bayesian} estimation for the four-parameter logistic model ({4PLM})},
  author={Zheng, Chanjin and Guo, Shaoyang and Kern, Justin L},
  journal={SAGE Open},
  volume={11},
  number={4},
  pages={21582440211052556},
  year={2021},
  publisher={SAGE Publications Sage CA: Los Angeles, CA},
  doi={10.1177/21582440211052556}
}
%----------------------------------------------------------------------------------------------
\clearpage
\appendix
\section{Appendix}\label{appendix}

%-----------------
\subsection{Proof of model equivalence}\label{sec:formal}
%-----------------

To prove the equivalence of the proposed 4P FA model defined in Section~\ref{model_spec} and the 4P IRT model (see Section~\ref{4P_IRT}), we first show the equivalence of their conditional probabilities. 
In the next theorem~\ref{sec:formal:marginal}, we show the equivalence of unconditional (marginal) probabilities $\textsf{P}(\mathbf{Y} = \mathbf{y})$ under the 4P FA model and the 4P IRT model. This result extends the work of \cite{takane1987relationship}, who demonstrated such equivalence under the 2P case. Establishing this equivalence becomes more intricate in the FA framework when incorporating the hierarchical structure of the 4P model. 
The equivalence of 3P FA and 3P IRT models follows as a special case by setting all inattention parameters to one. Before proceeding, we introduce technical lemmas that will be used in the proofs.
While a more general version of this proof has been provided in the context of knowledge space theory \cite{noventa2019generalization}, we present a direct proof using FA and IRT structures and notations requiring no background in alternative theoretical frameworks.

\begin{lemma}\label{A1}
Let observed response $\mathbf{Y}$ and latent response $\mathbf{Y}^*$ follow the dichotomized 2P FA defined by Equations~\eqref{eq:FA1} and~\eqref{eq:FA2}, where elements of error terms $\boldsymbol{\epsilon}$ follow a centered logistic distribution (2PL FA model) or normal distribution (2PNO FA), respectively. Then, the conditional cumulative probability of the latent item performance has the form
\begin{align*}
    \textsf{P}(Y^*_{i} \geq \tau_{i} | \theta) = F\left(\frac{\alpha_{i}\theta - \tau_{i}}{u_{i}}\right),
\end{align*}
where $F()$ is the cumulative distribution function of the standard logistic or normal distribution, respectively, depending on the distribution of $\boldsymbol{\epsilon}$.
\end{lemma}
\textbf{Proof:} 
For the 2PNO FA, the elements of $\boldsymbol{\epsilon}$ follow the Normal distribution $\epsilon_{i} \sim \textsf{N}(0,u^2_{i})$, therefore $Y^*_{i} | \theta \sim \textsf{N}(\alpha_{i}\theta,u^2_{i})$ and, using substitution $z = \frac{\alpha_{i}\theta - y}{u_{i}}$, then $dz = -\frac{1}{u_{i}}dy$, and adjusting the limits of integral accordingly, we get: 
    \begin{align*}
        \textsf{P}(Y^*_{i} \geq \tau_{i}|\theta) = \int_{\tau_{i}}^{\infty} \frac{1}{\sqrt{2\pi}\cdot u_{i}}e^{-\frac{1}{2u^2_{i}}(\alpha_{i}\theta - y)^2}dy = \int_{-\infty}^{\frac{\alpha_{i}\theta - \tau_{i}}{u_{i}}} \frac{1}{\sqrt{2\pi}}e^{-\frac{1}{2}z^2}dy=F\left(\frac{\alpha_{i}\theta - \tau_{i}}{u_{i}}\right).
    \end{align*}

For the 2PL FA, the elements of error terms $\boldsymbol{\epsilon}$ follow the logistic distribution $\epsilon_{i} \sim logist(0,u^2_{i})$. We consider $\epsilon^{std}_{i}\sim logist(0,1)$ the standard logistic distribution with mean $0$, variance $\frac{\pi^2}{3}$ and CDF $\frac{e^x}{1 + e^x}$. We have 
    \begin{align*}
        Y^*_{i} = \alpha_{i}\theta+\epsilon_{i},
    \text{   or equivalently   }\;\;\;
        Y^*_{i} = \alpha_{i}\theta + u_{i} \cdot \epsilon^{std}_{i}.
    \end{align*}
    Denote
    \begin{align*}
        F(x) = \textsf{P}(\epsilon^{std}_{i}\leq x) = \frac{e^x}{1 + e^x} = \frac{1}{1+e^{-x}}.
    \end{align*}
Then, by absolute continuity, and independence from $\theta$, we get
    \begin{align*}
        \textsf{P}(Y^*_{i} \geq \tau_{i} | \theta)& = 1 - \textsf{P}(Y^*_{i} < \tau_{i} | \theta) = 1 - \textsf{P}(\alpha_{i}\theta + u_{i} \cdot \epsilon^{std}_{i} < \tau_{i} | \theta)\\
        & = 1 - \textsf{P}\left(\epsilon^{std}_{i} < \frac{\tau_{i} - \alpha_{i}\theta}{u_{i}}\Big | \theta\right) = 1 - \int_{-\infty}^{\frac{\tau_{i}-\alpha_{i}\theta}{u_{i}}} f(\epsilon^{std}_{i} | \theta)d\epsilon^{std}_{i} \\
        & = 1 - \int_{-\infty}^{\frac{\tau_{i}-\alpha_{i}\theta}{u_{i}}} \frac{f(\epsilon^{std}_{i} , \theta)}{f(\theta)} d\epsilon^{std}_{i} 
        = 1 - \int_{-\infty}^{\frac{\tau_{i}-\alpha_{i}\theta}{u_{i}}} \frac{f(\epsilon^{std}_{i})f(\theta)}{f(\theta)} d\epsilon^{std}_{i} \\
        & = 1 - \int_{-\infty}^{\frac{\tau_{i}-\alpha_{i}\theta}{u_{i}}} f(\epsilon^{std}_{i}) d\epsilon^{std}_{i} 
        = 1 - \textsf{P}\left(\epsilon^{std}_{i} \leq \frac{\tau_{i} - \alpha_{i}\theta}{u_{i}}\right) \\
        & = 1 - \frac{1}{1 + e^{-(\tau_{i} - \alpha_{i}\theta)/u_{i}}}\\
        & = \frac{e^{(\alpha_{i}\theta - \tau_{i})/u_{i}}}{1 + e^{(\alpha_{i}\theta - \tau_{i})/u_{i}}} = F\left(\frac{\alpha_{i}\theta - \tau_{i}}{u_{i}}\right).
    \end{align*}
Alternatively, we could directly use $Y^*_{i} | \theta \sim logist(\alpha_{i}\theta,u^2_{i})$.

%----------------------------------------------------------------

\begin{lemma}\label{A2} 
Let $m \in \mathbb{N}$ and $\mathbf{y} = (y_1,\dots,y_m)^T\in\{0,1\}^m$. Define $x_{i} := \frac{\alpha_{i}\theta - \tau_{i}}{u_{i}}$ for brevity. Then, with 
$\mathbf{Z}$ defined as in~\eqref{eq:Zi}, we can exchange summation and product as:
    \begin{align*}
        &\sum_{\mathbf{z}\in \{0,1\}^m}\left\{ \prod_{i = 1}^m \left[ 
        \textsf{P}(Y_{i} = y_i | Z_{i} = z_i) \cdot F(x_{i})^{z_i}\left(1 - F(x_{i})\right)^{1 - z_i}
        \right]\right\}\\
        & = \prod_{i = 1}^m \left\{ \sum_{z_i \in \{0,1\}}\left[ 
        \textsf{P}(Y_{i} = y_i | Z_{i} = z_i) \cdot F(x_{i})^{z_i}\left(1 - F(x_{i})\right)^{1 - z_i}
        \right]\right\}\\
        & = \prod_{i = 1}^m \Big[\textsf{P}(Y_{i} = y_i | Z_{i} = 0)\left(1 - F(x_{i})\right) + \textsf{P}(Y_{i} = y_i | Z_{i} = 1)F(x_{i}) \Big]
    \end{align*}
\end{lemma}
\textbf{Proof by induction:}\\
We use $\mathbf{z}_{(-m)}$ to denote $\mathbf{z}$ without the $m$-th element. For $m = 1$, the formula holds trivially by direct evaluation:
\begin{align*}
    &\sum_{z_1 \in \{0,1\}}\left\{ 
        \textsf{P}(Y_{1} = y_1 | Z_{1} = z_1) \cdot F(x_{1})^{z_1}\left(1 - F(x_{1})\right)^{1 - z_1} \right\} \\
        & = \prod_{i = 1}^1 \Big[\textsf{P}(Y_{i} = y_i | Z_{i} = 0)\left(1 - F(x_{i})\right) + \textsf{P}(Y_{i} = y_i | Z_{i} = 1)F(x_{i}) \Big].
\end{align*}
\textbf{Inductive step:} Assume the formula holds for $m-1$. For $m \in \mathbb{N}, \; m > 1$, we split over $z_m \in \{0,1\}$ and factor out terms common across summation as shown: 
        \begin{align*}
        &\sum_{\mathbf{z} \in \{0,1\}^m}\left\{ \prod_{i = 1}^m \left[ 
        \textsf{P}(Y_{i} = y_i | Z_{i} = z_i) \cdot F(x_{i})^{z_i}\left(1 - F(x_{i})\right)^{1 - z_i}
        \right]\right\}
        \end{align*}
        %We split this sum according to the value of the last term $z_m$ of $\mathbf{z}$, which can be only $0$ or $1$:
        \begin{align*}
        &= \sum_{\substack{\mathbf{z}_{(-m)}\in \{0,1\}^{m - 1}\\
                  z_m = 0}}\left\{ \prod_{i = 1}^m \left[ 
        \textsf{P}(Y_{i} = y_i | Z_{i} = z_i) \cdot F(x_{i})^{z_i}\left(1 - F(x_{i})\right)^{1 - z_i}
        \right]\right\}\\
        &+ \sum_{\substack{\mathbf{z}_{(-m)}\in \{0,1\}^{m - 1}\\
                  z_m = 1}}\left\{ \prod_{i = 1}^m \left[ 
        \textsf{P}(Y_{i} = y_i | Z_{i} = z_i) \cdot F(x_{i})^{z_i}\left(1 - F(x_{i})\right)^{1 - z_i}
        \right]\right\}
        \end{align*}
        %In each sum separately, all products have the same term when $i=m$ as $z_m$ is fixed. We put this term in front of each sum:
        \begin{align*}
        %& 
        = \textsf{P}(Y_{m} = y_m | Z_{m} = 0) \left(1 - F(x_{m})\right)%\\
        %& 
        \sum_{\mathbf{z}_{(-m)}\in \{0,1\}^{m - 1}}\left\{ \prod_{i = 1}^{m - 1} \left[ 
        \textsf{P}(Y_{i} = y_i | Z_{i} = z_i) \cdot F(x_{i})^{z_i}\left(1 - F(x_{i})\right)^{1 - z_i}
        \right]\right\}\\[1.5ex]
        %& 
        + \textsf{P}(Y_{m} = y_m | Z_{m} = 1) F(x_{m})%\\
        %& 
        \sum_{\mathbf{z}_{(-m)}\in \{0,1\}^{m - 1}}\left\{ \prod_{i = 1}^{m - 1} \left[ 
        \textsf{P}(Y_{i} = y_i | Z_{i} = z_i)\cdot F(x_{i})^{z_i}\left(1 - F(x_{i})\right)^{1 - z_i}
        \right]\right\}.
        \end{align*}
        The rest follows by applying the inductive hypothesis and reassembling the full product using the distributive property: %Now, each sum has only $m - 1$ terms, so we can use the induction step:
        \begin{align*}
        %&
        =\textsf{P}(Y_{m} = y_m | Z_{m} = 0) \left(1 - F(x_{m})\right)%\\
        & \cdot \prod_{i = 1}^{m - 1} \big[\textsf{P}(Y_{i} = y_i | Z_{i} = 0)\left(1 - F(x_{i})\right) + \textsf{P}(Y_{i} = y_i | Z_{i} = 1)F(x_{i}) \big]\\
        %& 
        + \textsf{P}(Y_{m} = y_m | Z_{m} = 1) F(x_{m})%\\
        & \cdot \prod_{i = 1}^{m - 1} \big[\textsf{P}(Y_{i} = y_i | Z_{i} = 0)\left(1 - F(x_{i})\right) + \textsf{P}(Y_{i} = y_i | Z_{i} = 1)F(x_{i}) \big]
        \end{align*}
        %Products are the same; therefore, after using the distributive property, we obtain the product of two terms:
        \begin{align*}
        %& 
        = \Big[\textsf{P}(Y_{m} = y_m | Z_{m} = 0) \left(1 - F(x_{m})\right) + \textsf{P}(Y_{m} = y_m | Z_{m}=1) F(x_{m}) \Big]\\
        %& 
        \cdot \prod_{i = 1}^{m - 1} \Big[\textsf{P}(Y_{i} = y_i | Z_{i} = 0)\left(1 - F(x_{i})\right) + \textsf{P}(Y_{i} = y_i | Z_{i} = 1)F(x_{i}) \Big]
        \end{align*}
        %which gives the desired form by combining products together:
        \begin{align*}
        %& 
        = \prod_{i = 1}^m \Big[\textsf{P}(Y_{i} = y_i | Z_{i} = 0)\left(1 - F(x_{i})\right) + \textsf{P}(Y_{i} = y_i | Z_{i} = 1)F(x_{i}) \Big].
    \end{align*}
%\end{itemize}

\begin{lemma}\label{A3}
Let the 2P FA model, defined by~\eqref{eq:FA1} and~\eqref{eq:FA2} in Section~\ref{sec:FA-IRT}, hold. Then the latent responses conditioned on the latent trait $Y^*_{i}|\theta$ for $i = 1, \dots, m$ are independent.
\end{lemma}

\textbf{Proof:}
To distinguish arguments of different densities in this proof, we append an index indicating the corresponding distribution. We aim to show that conditional density $f_{\mathbf{Y}^*|\theta}(\mathbf{y}^*|\theta)$ can be written as the product of the marginal conditional densities $f_{Y_{i}^*|\theta}(y_{i}^*|\theta)$. To do this, we first derive the joint densities $f_{\mathbf{Y}^*,\theta}(\mathbf{y}^*,\theta)$ and $f_{Y_{i}^*,\theta}(y_i^*,\theta)$. 

The random variables $\theta$ and $\epsilon_{i}$ for $i=1,\dots,m$ are assumed to be independent and continuously distributed. Therefore, the joint density exists and takes the form: 
\begin{align*}
    f_{\mathbf{\epsilon},\theta}(\mathbf{\epsilon},\theta) = \prod_{i=1}^{m} f_{\epsilon_{i}}(\epsilon_{i}) \cdot f_{\theta}(\theta). 
\end{align*}
The random vector $(\mathbf{Y}^*,\theta)^T$ can be obtained from $(\mathbf{\epsilon},\theta)^T$ through the  transformation $H$:
\begin{align*}
\begin{pmatrix}
    Y_{1}^* \\
    \vdots \\
    Y_{m}^* \\
    \theta
\end{pmatrix}  
= 
H\begin{pmatrix}
    \epsilon_{1} \\
    \vdots \\
    \epsilon_{m} \\
    \theta
\end{pmatrix}  
:=
\begin{pmatrix}
    \alpha_{1}\theta + \epsilon_{1} \\
    \vdots \\
    \alpha_{m}\theta + \epsilon_{m} \\
    \theta
\end{pmatrix}.  
\end{align*}
Using the transformation theorem \parencite[see][Theorem 4.4.3]{poirier1995intermediate}, the joint density becomes:
\begin{align}\label{lem21}
    f_{\mathbf{Y}^*,\theta}(\mathbf{y}^*,\theta) = f_{\mathbf{\epsilon},\theta}(\mathbf{y}^*-\mathbb{A}\theta,\theta) = \prod_{i=1}^{m} f_{\epsilon_{i}}(y_{i}^*-\alpha_{i}\theta) \cdot f_{\theta}(\theta).
\end{align}
Similarly, for each $i$, we obtain: 
\begin{align}\label{lem22}
    f_{Y_{i}^*,\theta}(y_i^*,\theta) = f_{\epsilon_{i}}(y_{i}^*-\alpha_{i}\theta) \cdot f_{\theta}(\theta),
    \hspace{2mm}\text{ for } i=1,\dots,m.
\end{align}
Now, using equations~\eqref{lem21} and~\eqref{lem22}, the conditional density becomes:
\begin{align*}
    f_{\mathbf{Y}^*|\theta}(\mathbf{y}^*|\theta) 
    &= 
    \frac{f_{\mathbf{Y}^*,\theta}(\mathbf{y}^*,\theta)}{f_{\theta}(\theta)} 
    = 
    \frac{\prod_{i=1}^{m} f_{\epsilon_{i}}(y_{i}^*-\alpha_{i}\theta) \cdot f_{\theta}(\theta)}{f_{\theta}(\theta)}\\
    &\\
    &= 
    \prod_{i=1}^{m} f_{\epsilon_{i}}(y_{i}^*-\alpha_{i}\theta) 
    =
    \prod_{i=1}^{m} \frac{f_{Y_{i}^*,\theta}(y_i^*,\theta)}{f_{\theta}(\theta)}
    =
    \prod_{i=1}^{m} f_{Y_{i}^*|\theta}(y_i^*|\theta),
\end{align*}
which concludes the proof.

%---------
\begin{theorem}[Equivalence of Conditional Probabilities]\label{sec:formal:conditional}
Consider $\textbf{Y}$ following the 4P FA model defined by Equations~\eqref{eq:FA1}, \eqref{eq:Zi} and \eqref{eq:4P-FA}. Then the conditional probabilities $\textsf{P}(Y_{i} = 1 | \theta)$ of a correct response are given by
\begin{align*}%\label{heur1}
    \textsf{P}(Y_{i} = 1 | \theta) = c_{i} + (d_{i} - c_{i})\cdot F\big(a_{i}(\theta - b_{i})\big),
\end{align*}
which is identical to the response probability in the 4P IRT model, as defined in Equation~\eqref{eq:4P-IRT}.
\end{theorem}

\textbf{Proof}: Using the law of total probability \parencite[see][p.~153]{mendenhall2012introduction}, the conditional probability can be decomposed as follows:
\begin{align*}
    \textsf{P}(Y_{i} = 1|\theta) = \textsf{P}(Y_{i} = 1|\theta,Y^*_{i} < \tau_{i})\textsf{P}(Y^*_{i} < \tau_{i}|\theta) + \textsf{P}(Y_{i} = 1 | \theta,Y^*_{i} \geq \tau_{i})\textsf{P}(Y^*_{i} \geq \tau_{i} | \theta).
\end{align*}
The conditional probabilities $\textsf{P}(Y_{i} = 1|\theta,Y'_{i} < \tau_{i})$ and $\textsf{P}(Y_{i} = 1|\theta,Y'_{i} \geq \tau_{i})$ do not depend on $\theta$ and are equal to $c_{i}$ and $d_{i}$, respectively. Therefore, we can rewrite:
\begin{align}\label{FAcond1}
    \textsf{P}(Y_{i} = 1|\theta) = c_{i} \cdot \textsf{P}(Y^*_{i} < \tau_{i}|\theta) + d_{i} \cdot\textsf{P}(Y^*_{i}\geq \tau_{i}|\theta).
\end{align}
According to Lemma~\ref{A1}, we can use the cumulative distribution function to replace probabilities:
\begin{align}\label{FAcond2}
    \textsf{P}(Y^*_{i} \geq \tau_{i}|\theta) = F\left(\frac{\alpha_{i}\theta - \tau_{i}}{u_{i}}\right)
 \text{ and } \textsf{P}(Y^*_{i} < \tau_{i}|\theta) = 1 - F\left(\frac{\alpha_{i}\theta - \tau_{i}}{u_{i}}\right).
\end{align}
Substituting \eqref{FAcond2} into \eqref{FAcond1}, we obtain:
\begin{align}\label{heur2}
    \textsf{P}(Y_{i} = 1 | \theta) & = c_{i}\left(1 - F\left(\frac{\alpha_{i}\theta - \tau_{i}}{u_{i}}\right)\right) + d_{i} F\left(\frac{\alpha_{i}\theta - \tau_{i}}{u_{i}}\right)\nonumber\\[1.5ex]
    & = c_{i} + (d_{i} - c_{i}) F\left(\frac{\alpha_{i}\theta - \tau_{i}}{u_{i}}\right).
\end{align}
Finally, the expression~\eqref{heur2} matches the form of the 4P IRT model in Equation~\eqref{eq:4P-IRT}, up to the parameter transformation defined in~\eqref{eq:eqvi}, which concludes the proof.

\vspace{1em}
%---------

\begin{theorem}[Equivalence of marginal probabilities]\label{sec:formal:marginal}
Consider $\textbf{Y}$ following the 4P FA model defined by Equations~\eqref{eq:FA1}, \eqref{eq:Zi} and \eqref{eq:4P-FA}. Then the marginal probabilities $\textsf{P}(Y_{i} = 1)$ of the correct responses in the 4P FA model match the marginal probabilities in the 4P IRT model.
\end{theorem}

\textbf{Proof}: The 4P IRT model is defined by Equations~\eqref{eq:4P-IRT} and \eqref{2}. Under this model, the marginal probability of observing a response pattern $\mathbf{y}$ is given by:
\begin{align}\label{irtform}
    \textsf{P}(\mathbf{Y} = \mathbf{y}) = \int_\Theta \prod_{i = 1}^m \Big[&\Big(c_{i} + (d_{i} - c_{i}) F\big(a_{i}(\theta - b_{i})\big)\Big)^{y_i}\nonumber\\
    &\Big(1 - c_{i} - (d_{i} - c_{i}) F\big(a_{i}(\theta - b_{i})\big)\Big)^{1 - y_i} \Big] f(\theta) d\theta.
\end{align}

Now consider $\mathbf{Y}$ following the 4P FA model as defined by Equations~\eqref{eq:FA1}, \eqref{eq:Zi}, and \eqref{eq:4P-FA}. 
Let the binary latent responses $Z_{i}$ be defined as in~\eqref{eq:Zi}, and define the regions $R_i^{z_i}$ associated with each $Z_i = z_i$:
\begin{align*}
    Z_{i} = \begin{cases}
        0 \; \; \; \; Y^*_{i} < \tau_{i} \; &||  \; \; R_{i}^0 = (-\infty, \tau_{i})\\
        1 \; \; \; \; Y^*_{i} \geq \tau_{i} \; &|| \; \; R_{i}^1 = \hspace{0.08cm}[\tau_{i}\phantom{,,},\infty).
       \end{cases}
\end{align*}
Define the Cartesian region $R^{\boldsymbol{z}} = R_{1}^{z_1}\times \dots \times R_{m}^{z_m}$. The marginal probability of observing response vector $\mathbf{y}$ is the sum over all possible latent patterns $\mathbf{z} \in \{0,1\}^m$: %of the conditional probabilities:
\begin{align*}
    \textsf{P}(\mathbf{Y} = \mathbf{y}) = \sum_{\mathbf{z}\in \{0,1\}^m}\Big\{ \textsf{P}(\mathbf{Y} = \mathbf{y} | \mathbf{Z} = \mathbf{z})\cdot  \textsf{P}(\mathbf{Z} = \mathbf{z})\Big\}.
\end{align*}
Due to item-level independence, the conditional probability factorizes: %while the term $\textsf{P}(\mathbf{Z} = \mathbf{z})$ is the integral of density $f(\mathbf{y}^*)$ of $\mathbf{Y}^*$ over corresponding regions:
\begin{align*}
    \textsf{P}(\mathbf{Y} = \mathbf{y}) = \sum_{\mathbf{z}\in \{0,1\}^m}\left\{ \prod_{i = 1}^m \big[ \textsf{P}(Y_{i} = y_i | Z_{i} = z_i)\big]\cdot  \int_{R^{\boldsymbol{z}}} f(\mathbf{y}^*)d\mathbf{y}^*\right\}.
\end{align*}
Each $(Y_{i},Z_{i})^T$ pair admits only four possible combinations. 
Table~\ref{probyz} summarizes the corresponding conditional probabilities, derived from Equation~\eqref{eq:4P-FA}.
\begin{table}[ht]
\centering
\begin{tabular}{r|c|c}
$\textsf{P}(Y_{i} | Z_{i})$&$Y_{i} = 0$ & $Y_{i} = 1$ \\
\hline
$Z_{i} = 0$ &$1 - c_{i}$ & $c_{i}$ \\
\hline
$Z_{i} = 1$ & $1 - d_{i}$ & $d_{i}$ \\
\end{tabular}
\caption{Conditional probabilities $\textsf{P}(Y_{i} = y_i | Z_{i} = z_i)$.}
\label{probyz}
\end{table}\\
Next, we express the marginal density $ f(\mathbf{y}^*)$ via the joint distribution of $(\mathbf{y}^*, \theta)$: 
$$ f(\mathbf{y}^*) = \int_{\Theta} f(\mathbf{y}^*, \theta) d\theta = \int_{\Theta} F'(\mathbf{y}^* | \theta)\cdot f(\theta) d\theta,$$
%We also use the fact that joint density $f(\mathbf{y}^*, \theta)$ can be represented as the product of conditional and marginal densities $F'(\mathbf{y}^* | \theta)\cdot f(\theta)$. Density $F'()$ corresponds to cumulative distribution function $F()$ from Equation~\eqref{eq:4P-IRT}, thus 
where $F'(\mathbf{y}^*|\theta)$ denotes the conditional density corresponding to the cumulative distribution function $F$ in Equation~\eqref{eq:4P-IRT}. Substituting into the previous expression:
\begin{align*}
    \textsf{P}(\mathbf{Y} = \mathbf{y}) = \sum_{\mathbf{z}\in \{0,1\}^m}\left\{ \prod_{i = 1}^m \big[ \textsf{P}(Y_{i} = y_i | Z_{i} = z_i)\big] \cdot  \int_{R^{\boldsymbol{z}}} \int_{\Theta} F'(\mathbf{y}^* | \theta)\cdot f(\theta) d\theta \, d\mathbf{y}^*\right\}.
\end{align*}
Applying Fubini's theorem \parencite[see][Chapter 5.2]{cohn2013measure} to switch the order of integration:
\begin{align*}
    \textsf{P}(\mathbf{Y} = \mathbf{y}) = \sum_{\mathbf{z}\in \{0,1\}^m}\left\{ \prod_{i = 1}^m \big[ \textsf{P}(Y_{i} = y_i | Z_{i} = z_i)\big] \cdot \int_{\Theta} \int_{R^{\boldsymbol{z}}} F'(\mathbf{y}^* | \theta)  d\mathbf{y}^* \cdot f(\theta) d\theta \right\}.
\end{align*}
By Lemma~\ref{A3}, the conditional independence of $Y_i^* | \theta$ implies the factorization of the inner integral: 
$$\int_{R^z} F'(\mathbf{y}^* | \theta)  d\mathbf{y}^* = \prod_{i = 1}^m \int_{R_{i}^{z_i}} F'(y^*_i | \theta)  d y^*_i.$$
Therefore,
\begin{align*}
    \textsf{P}(\mathbf{Y} = \mathbf{y}) = \sum_{\mathbf{z}\in \{0,1\}^m}\left\{ \prod_{i = 1}^m \big[ \textsf{P}(Y_{i} = y_i | Z_{i} = z_i)\big]\cdot \int_{\Theta} \prod_{i = 1}^m \left[\int_{R_{i}^{z_i}} F'(y^*_i | \theta)  d y^*_i \right] \cdot f(\theta) d\theta \right\}.
\end{align*}
\newline
Since $R_{i}^{z_i}$ corresponds to the cumulative probabilities of $Y_i^*$ being above or below $\tau_i$, we can write: %are intervals that depend on values $z_i$ and $\tau_i$, and we can thus rewrite the integrals more conveniently as probabilities:
\begin{align*}
\int_{R_i^{z_i}} F'(y_i^* | \theta)  dy_i^* = \textsf{P}(Y_i^* \geq \tau_i | \theta)^{z_i} \cdot \textsf{P}(Y_i^* < \tau_i | \theta)^{1 - z_i}.
\end{align*}
Thus,
\begin{align*}
    \textsf{P}(\mathbf{Y} = \mathbf{y}) = \sum_{\mathbf{z}\in \{0,1\}^m}\left\{ \prod_{i = 1}^m \big[ \textsf{P}(Y_{i} = y_i | Z_{i} = z_i)\big]\cdot \int_{\Theta} \prod_{i = 1}^m \left[ \textsf{P}(Y^*_{i} \geq \tau_{i} | \theta)^{z_i}\; \textsf{P}(Y^*_{i} < \tau_{i} | \theta)^{1 - z_i} \right] \cdot f(\theta) d\theta \right\}.
\end{align*}
\newline
Using Lemma~\ref{A1} to rewrite $\textsf{P}(Y^*_{i} \geq \tau_{i} | \theta)$ and $\textsf{P}(Y'_{i} < \tau_{i} | \theta)$ as cumulative distribution functions:
\begin{align*}
    = \sum_{\mathbf{z} \in \{0,1\}^m}\left\{ \prod_{i = 1}^m \big[ \textsf{P}(Y_{i} = y_i | Z_{i} = z_i)\big]\cdot \int_{\Theta} \prod_{i = 1}^m \left[ F\left(\frac{\alpha_{i}\theta - \tau_{i}}{u_{i}}\right)^{z_i} \left(1 - F\left(\frac{\alpha_{i}\theta - \tau_{i}}{u_{i}}\right) \right)^{1 - z_i} \right] \cdot f(\theta) d\theta \right\}.
\end{align*}
\newline
We bring the constant (with respect to $\theta$) product term %$\prod_{i = 1}^m \big[ \textsf{P}(Y_{i} = y_i | Z_{i} = z_i)\big]$ 
inside the integral and combine products:

\begin{align*}
  =  \sum_{\mathbf{z} \in \{0,1\}^m}\left\{ \int_{\Theta} \prod_{i = 1}^m \left[ \textsf{P}(Y_{i} = y_i | Z_{i} = z_i) \cdot F\left(\frac{\alpha_{i}\theta - \tau_{i}}{u_{i}}\right)^{z_i} \left(1 - F\left(\frac{\alpha_{i}\theta - \tau_{i}}{u_{i}}\right) \right)^{1 - z_i} \right] \cdot f(\theta) d\theta \right\}.
\end{align*}
\newline
Since the sum %$\sum_{\mathbf{z} \in \{0,1\}^m}$ 
is finite, we may interchange the sum and integral: 
%use the linearity of integral and write:
\begin{align*}
       \textsf{P}(\mathbf{Y} = \mathbf{y}) = \int_{\Theta} \sum_{\mathbf{z}\in \{0,1\}^m}\prod_{i = 1}^m \left[ \textsf{P}(Y_{i} = y_i | Z_{i} = z_i) 
       \cdot F\left(\frac{\alpha_{i}\theta - \tau_{i}}{u_{i}}\right)^{z_i} \left(1 - F\left(\frac{\alpha_{i}\theta - \tau_{i}}{u_{i}}\right) \right)^{1 - z_i} \right] \cdot f(\theta) d\theta .
\end{align*}
\newline
Applying Lemma~\ref{A2}, we can exchange the sum-product order, %: $\sum_{\mathbf{z}\in \{0,1\}^m}\prod_{i = 1}^m = \prod_{i = 1}^m \sum_{z_i \in \{0,1\}}$,
%. Breaking down the sum, we obtain:
yielding: 
\begin{align*}
\textsf{P}(\mathbf{Y} = \mathbf{y}) = \int_{\Theta} \prod_{i = 1}^m \left[ \sum_{z_i \in {0,1}} \textsf{P}(Y_i = y_i \mid Z_i = z_i) \cdot F\left(\frac{\alpha_i \theta - \tau_i}{u_i} \right)^{z_i} \left(1 - F\left(\frac{\alpha_i \theta - \tau_i}{u_i} \right) \right)^{1 - z_i} \right] f(\theta)d\theta,
\end{align*}
and by evaluating for $z_i$:
\begin{align*}
       \textsf{P}(\mathbf{Y} = \mathbf{y}) = \int_{\Theta}\prod_{i = 1}^m \Bigg[ &\textsf{P}(Y_{i} = y_i | Z_{i} = 0)\cdot  \left(1 - F\left(\frac{\alpha_{i}\theta - \tau_{i}}{u_{i}}\right) \right)\\
       & + \textsf{P}(Y_{i} = y_i | Z_{i} = 1)\cdot  F\left(\frac{\alpha_{i}\theta - \tau_{i}}{u_{i}}\right) \Bigg] \cdot f(\theta) d\theta .
\end{align*}
%\newline
Rewriting using the value of $Y_i$ in the exponent, we obtain:
%Evaluating the inner sums, we get: %This can be rewritten more conveniently, using the fact that $y_i$ can be only $0$ or $1$:
\begin{align*}
       \textsf{P}(\mathbf{Y} = \mathbf{y}) = \int_{\Theta}\prod_{i = 1}^m \Bigg[ &\Bigg\{\textsf{P}(Y_{i} = 1 | Z_{i} = 0)\cdot  \left(1 - F\left(\frac{\alpha_{i}\theta - \tau_{i}}{u_{i}}\right) \right)\\
       & + \textsf{P}(Y_{i} = 1 | Z_{i} = 1)\cdot  F\left(\frac{\alpha_{i}\theta - \tau_{i}}{u_{i}}\right)\Bigg\}^{y_i}\\
       &\cdot \Bigg\{\textsf{P}(Y_{i} = 0 | Z_{i} = 0)\cdot  \left(1 - F\left(\frac{\alpha_{i}\theta - \tau_{i}}{u_{i}}\right) \right)\\
       & + \textsf{P}(Y_{i} = 0 | Z_{i} = 1)\cdot  F\left(\frac{\alpha_{i}\theta - \tau_{i}}{u_{i}}\right)\Bigg\}^{1 - y_i} \Bigg] \cdot f(\theta) d\theta.
\end{align*}
%\newline
Further, using the conditional probabilities from~Table~\ref{probyz}, we get:
\begin{align*}
       \textsf{P}(\mathbf{Y} = \mathbf{y}) = \int_{\Theta}\prod_{i = 1}^m \Bigg[ &\Bigg\{c_{i}\cdot  \left(1 - F\left(\frac{\alpha_{i}\theta - \tau_{i}}{u_{i}}\right) \right)\\
       & + d_{i}\cdot  F\left(\frac{\alpha_{i}\theta - \tau_{i}}{u_{i}}\right)\Bigg\}^{y_i}\\
       &\cdot \Bigg\{(1 - c_{i})\cdot  \left(1 - F\left(\frac{\alpha_{i}\theta - \tau_{i}}{u_{i}}\right) \right)\\
       & + (1 - d_{i}) \cdot  F\left(\frac{\alpha_{i}\theta - \tau_{i}}{u_{i}}\right)\Bigg\}^{1 - y_i} \Bigg] \cdot f(\theta) d\theta .
\end{align*}
%\newline
Finally, after regrouping terms, we obtain the same form as in Equation~\eqref{irtform}.
\begin{align*}
    \textsf{P}(\mathbf{Y} = \mathbf{y}) = \int_\Theta \prod_{i = 1}^m \Big[&\left(c_{i} + (d_{i} - c_{i})F\left(\frac{\alpha_{i}\theta - \tau_{i}}{u_{i}}\right)\right)^{y_i}\nonumber\\
    &\cdot \left(1 - c_{i} - (d_{i} - c_{i})F\left(\frac{\alpha_{i}\theta - \tau_{i}}{u_{i}}\right)\right)^{1 - y_i} \Big] f(\theta) d\theta,
\end{align*}
which matches~\eqref{irtform} up to reparameterization~\eqref{eq:eqvi}. This completes the proof.

\clearpage

%-----------------
\subsection{Simulation study tables}\label{sec:sim_tables}
%-----------------
In this section, we provide tables with means and medians of the MSE from the simulation study.

\begin{table}[!ht]
    \centering
    \includegraphics[scale=0.62, angle=90]{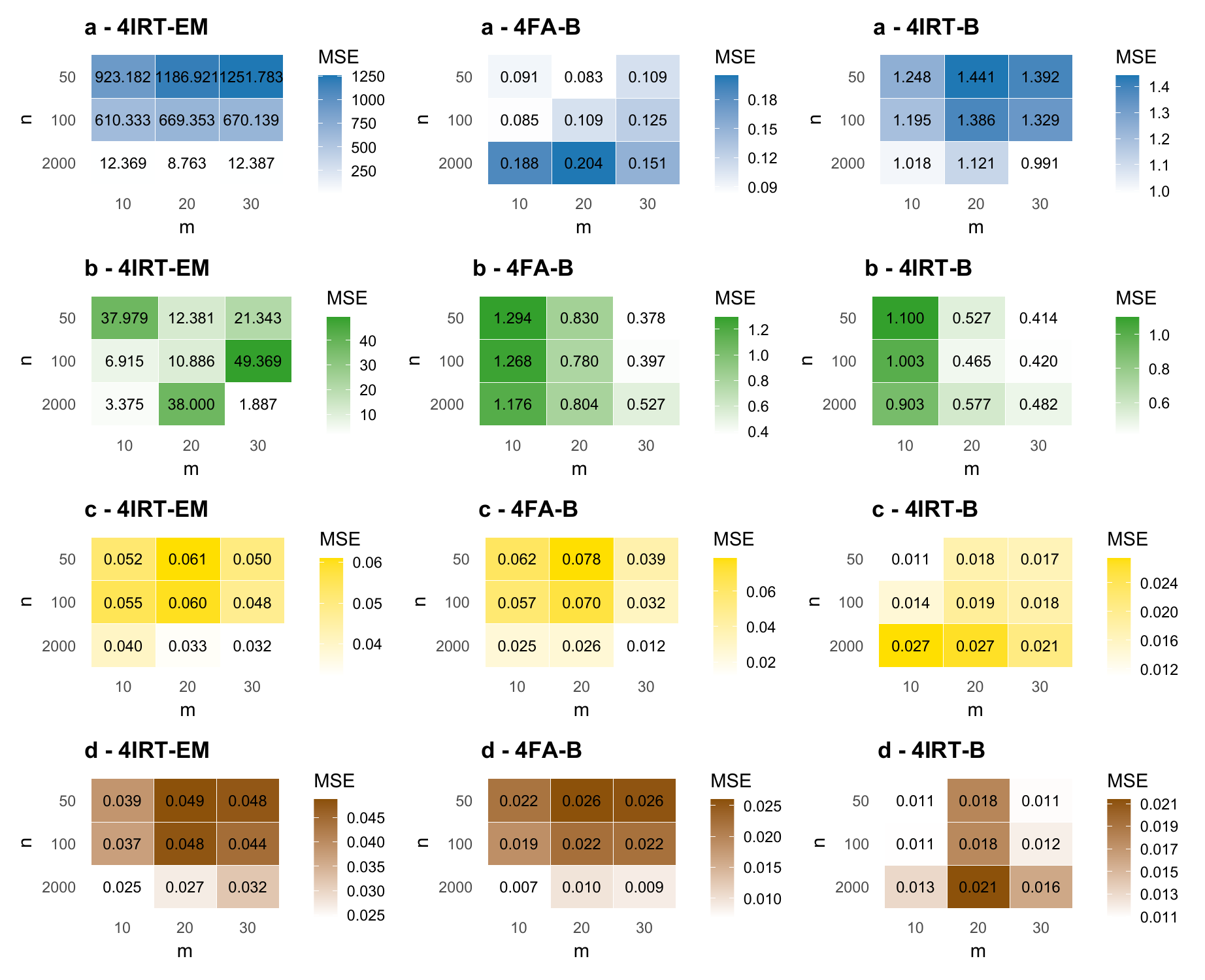}
     \vspace{-4mm}
    \caption{Comparison of means of Mean Square Errors (MSE) for each scenario and parameter discrimination (a), difficulty (b), guessing (c), and inattention (d) from simulation study. Left side (4IRT-EM) for IRT model fitted by EM algorithm using mirt package, middle (4FA-B) is for 4PL FA model using Bayesian estimates from JAGS, and right side (4IRT-B) is for typical 4PL IRT parameterization using Bayesian estimates from JAGS. Estimated values are compared to true simulated data.}
    \label{tab_sim_mean}
\end{table}

\begin{table}[!ht]
    \centering
    \includegraphics[scale=0.62, angle=90]{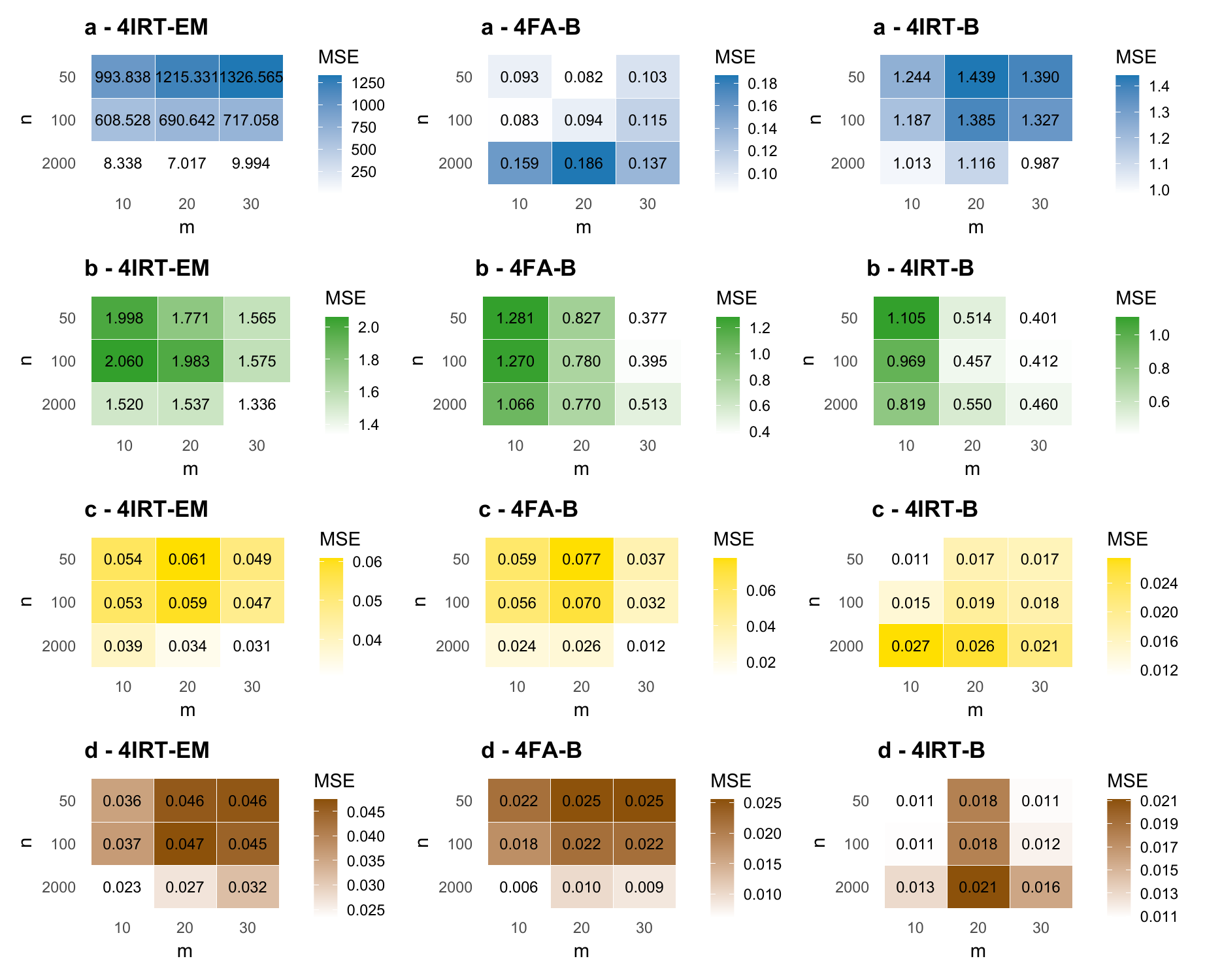}
     \vspace{-4mm}
    \caption{Comparison of medians of Mean Square Errors (MSE) for each scenario and parameter discrimination (a), difficulty (b), guessing (c), and inattention (d) from simulation study. Left side (4IRT-EM) for IRT model fitted by EM algorithm using mirt package, middle (4FA-B) is for 4PL FA model using Bayesian estimates from JAGS, and right side (4IRT-B) is for typical 4PL IRT parameterization using Bayesian estimates from JAGS. Estimated values are compared to true simulated data.}
    \label{tab_sim_median}
\end{table}

\clearpage

%-----------------
\subsection{Real data tables}\label{sec:tabs}
%-----------------

This section contains summary tables of item parameter estimates for the MSATB dataset (Table~\ref{tab1}) and the Anxiety dataset (Table~\ref{tab2}), provided to demonstrate their numerical similarity across estimation methods. Table~\ref{tabNGNI3} presents the averaged NGNI total scores, which are unique to each response pattern. A corresponding table based on the Anxiety dataset and results from \texttt{pymc} implementation is available in the Electronic Supplementary Material. %ed R-scripts and Jupyter notebooks.
Finally, we include Tables~\ref{mse_mirt_tab_1} and~\ref{mse_mirt_tab_2}, which report mean square error comparisons for estimated IRT parameters based on real data.
\newpage

\begin{sideways}
\centering
\begin{minipage}{1.5\textwidth}
\captionsetup[table]{width=0.95\linewidth}
\captionsetup{type=table}
\centering
\vspace{2cm}
\begin{tabular}{r|rrrr|rrrr|rrrr|rrrr}
\hline
&\multicolumn{4}{c|}{4FA-B-JAGS}&\multicolumn{4}{c|}{4FA-B-PyMC}&\multicolumn{4}{c}{4-IRT-B-JAGS}&\multicolumn{4}{c}{4-IRT-EM-mirt}\\
& $a$ & $b$ & $c$ & $d$ & $a$ & $b$ & $c$ & $d$ & $a$ & $b$ & $c$ & $d$ & $a$ & $b$ & $c$ & $d$ \\
\hline
Item49 & 3.0534 & -0.0872 & 0.7099 & 0.9883 & 2.8880 & -0.1310 & 0.7000 & 0.9900 & 1.8875 & -0.6526 & 0.5691 & 0.9850 & 2.9137 & -0.0030 & 0.7124 & 0.9991 \\
Item27 & 1.5304 & 1.0326 & 0.0533 & 0.8271 & 1.5190 & 1.0570 & 0.0530 & 0.8350 & 1.8488 & 1.0172 & 0.0744 & 0.8153 & 1.3717 & 1.3083 & 0.0481 & 0.9927 \\
Item41 & 3.8555 & 0.6138 & 0.1470 & 0.9717 & 3.9490 & 0.6180 & 0.1480 & 0.9710 & 2.7563 & 0.5688 & 0.1157 & 0.9699 & 4.0669 & 0.6359 & 0.1503 & 0.9832 \\
Item7 & 1.5632 & 0.0101 & 0.0314 & 0.8057 & 1.3900 & 0.1600 & 0.0360 & 0.8630 & 1.7503 & 0.0851 & 0.0664 & 0.8094 & 1.3610 & 0.0585 & 0.0004 & 0.8617 \\
Item38 & 2.2494 & -0.2776 & 0.2854 & 0.9885 & 2.3280 & -0.2530 & 0.2970 & 0.9880 & 2.2358 & -0.3469 & 0.2558 & 0.9805 & 2.2342 & -0.2252 & 0.2906 & 0.9989 \\
Item28 & 2.4675 & 1.3910 & 0.0972 & 0.7736 & 2.4090 & 1.4030 & 0.0960 & 0.7780 & 2.2464 & 1.4845 & 0.0941 & 0.8404 & 2.3462 & 1.4463 & 0.0958 & 0.8204 \\
Item9 & 2.0934 & -0.8553 & 0.1571 & 0.8929 & 2.0670 & -0.8370 & 0.1620 & 0.8940 & 2.0351 & -0.8024 & 0.1791 & 0.8987 & 1.7721 & -1.0648 & 0.0138 & 0.9003 \\
Item47 & 1.6244 & -0.9960 & 0.7195 & 0.9932 & 1.5890 & -1.0840 & 0.7010 & 0.9930 & 1.6600 & -1.9420 & 0.3368 & 0.9823 & 1.5003 & -0.9963 & 0.7023 & 0.9996 \\
Item75 & 2.4810 & 0.3508 & 0.3100 & 0.9384 & 2.3120 & 0.3520 & 0.3030 & 0.9470 & 2.2126 & 0.3038 & 0.2861 & 0.9388 & 2.7454 & 0.3666 & 0.3163 & 0.9307 \\
Item17 & 1.8245 & 1.0735 & 0.1012 & 0.9161 & 1.8830 & 1.0770 & 0.1040 & 0.9120 & 2.0777 & 1.0144 & 0.1104 & 0.8770 & 1.6874 & 1.1989 & 0.0965 & 0.9968 \\
Item76 & 3.1441 & 0.4481 & 0.3374 & 0.9765 & 3.1250 & 0.4600 & 0.3380 & 0.9790 & 2.4272 & 0.3599 & 0.2960 & 0.9727 & 2.9474 & 0.4995 & 0.3336 & 0.9990 \\
Item10 & 2.3597 & -0.8978 & 0.3512 & 0.9591 & 2.0560 & -1.0310 & 0.2800 & 0.9630 & 2.0565 & -1.1055 & 0.2285 & 0.9606 & 2.7879 & -0.8121 & 0.3794 & 0.9562 \\
Item64 & 1.4334 & 1.0061 & 0.3368 & 0.9271 & 1.4280 & 0.9530 & 0.3290 & 0.9170 & 1.6723 & 0.8218 & 0.3327 & 0.8719 & 1.6399 & 1.2306 & 0.3566 & 0.9991 \\
Item45 & 1.4824 & -0.3764 & 0.2109 & 0.9819 & 1.5910 & -0.2630 & 0.2580 & 0.9830 & 1.6646 & -0.3676 & 0.2292 & 0.9632 & 1.4947 & -0.2370 & 0.2423 & 0.9998 \\
Item24 & 1.3382 & 0.3914 & 0.0195 & 0.8738 & 1.3040 & 0.4510 & 0.0200 & 0.8950 & 1.5954 & 0.3766 & 0.0527 & 0.8393 & 1.1597 & 0.5597 & 0.0003 & 0.9767 \\
Item1 & 1.9906 & 0.6627 & 0.5378 & 0.9550 & 1.9460 & 0.6430 & 0.5360 & 0.9520 & 2.0187 & 0.5570 & 0.5261 & 0.9383 & 2.7463 & 0.7441 & 0.5619 & 0.9539 \\
Item68 & 1.2251 & -0.7731 & 0.1000 & 0.9273 & 1.2490 & -0.7030 & 0.1330 & 0.9300 & 1.6396 & -0.7228 & 0.1592 & 0.8859 & 1.0419 & -0.9068 & 0.0016 & 0.9632 \\
Item61 & 1.5039 & -0.6412 & 0.3276 & 0.9717 & 1.5150 & -0.6690 & 0.3140 & 0.9740 & 1.6684 & -0.8147 & 0.2449 & 0.9568 & 1.6246 & -0.4938 & 0.3625 & 0.9798 \\
Item25 & 1.1660 & 0.0716 & 0.2818 & 0.9537 & 1.1940 & 0.0740 & 0.2890 & 0.9530 & 1.6486 & 0.1297 & 0.3434 & 0.9050 & 1.1788 & 0.3547 & 0.3161 & 0.9987 \\
Item2 & 3.1961 & 1.2081 & 0.1634 & 0.8910 & 2.9300 & 1.2380 & 0.1610 & 0.9050 & 2.5049 & 1.2438 & 0.1523 & 0.9105 & 2.6586 & 1.3442 & 0.1602 & 0.9983 \\
\hline
\end{tabular}
\captionof{table}{Comparison of estimated item parameters for the MSATB dataset using four methods: (1) the 4PL FA model estimated via Bayesian methods in \texttt{JAGS} and (2) \texttt{pymc}, and (3) 4PL IRT model with typical IRT parameterization via Bayesian methods in \texttt{JAGS}, and (4) the 4PL IRT model estimated via the EM algorithm in the \texttt{mirt} package.}
\label{tab1}
\end{minipage}
\end{sideways}

\begin{sidewaystable}[htbp]
\centering
\begin{tabular}{r|rrrr|rrrr|rrrr|rrrr}
\hline
&\multicolumn{4}{c|}{4FA-B-JAGS}&\multicolumn{4}{c|}{4FA-B-PyMC}&\multicolumn{4}{c}{4-IRT-B-JAGS}&\multicolumn{4}{c}{4-IRT-EM-mirt}\\
& $a$ & $b$ & $c$ & $d$ & $a$ & $b$ & $c$ & $d$ & $a$ & $b$ & $c$ & $d$ & $a$ & $b$ & $c$ & $d$ \\
\hline
R1 & 3.5049 & 0.4832 & 0.0034 & 0.9880 & 3.4930 & 0.4840 & 0.0040 & 0.9870 & 2.9253 & 0.5402 & 0.0082 & 0.9766 & 3.6613 & 0.4721 & 0.0000 & 1.0000 \\
R2 & 3.2431 & 0.5606 & 0.0045 & 0.9817 & 3.2460 & 0.5640 & 0.0050 & 0.9820 & 2.7637 & 0.6284 & 0.0094 & 0.9683 & 3.3315 & 0.5521 & 0.0000 & 1.0000 \\
R3 & 3.7780 & 0.6606 & 0.0060 & 0.9850 & 3.7750 & 0.6630 & 0.0070 & 0.9860 & 3.0309 & 0.7555 & 0.0095 & 0.9760 & 3.8107 & 0.6359 & 0.0000 & 0.9999 \\
R4 & 3.3936 & -0.0522 & 0.0087 & 0.9946 & 3.3800 & -0.0500 & 0.0090 & 0.9940 & 2.8788 & -0.0766 & 0.0162 & 0.9880 & 3.5536 & -0.0430 & 0.0000 & 1.0000 \\
R5 & 3.2956 & 0.7169 & 0.0039 & 0.9833 & 3.2880 & 0.7190 & 0.0040 & 0.9830 & 2.7959 & 0.8089 & 0.0083 & 0.9693 & 3.4227 & 0.6930 & 0.0000 & 0.9999 \\
R6 & 2.9339 & 0.5373 & 0.0351 & 0.9890 & 2.9330 & 0.5420 & 0.0370 & 0.9890 & 2.5407 & 0.5916 & 0.0367 & 0.9760 & 3.1368 & 0.5255 & 0.0346 & 1.0000 \\
R7 & 3.3773 & -0.2411 & 0.0636 & 0.9959 & 3.3410 & -0.2480 & 0.0600 & 0.9950 & 2.7859 & -0.3185 & 0.0523 & 0.9904 & 3.5484 & -0.2220 & 0.0541 & 1.0000 \\
R8 & 4.6873 & 0.4916 & 0.1057 & 0.8418 & 4.4430 & 0.4950 & 0.1050 & 0.8460 & 2.6312 & 0.5929 & 0.0855 & 0.8850 & 5.7208 & 0.4622 & 0.1100 & 0.8362 \\
R9 & 1.8268 & 0.2391 & 0.0137 & 0.9549 & 1.8140 & 0.2440 & 0.0140 & 0.9560 & 1.7970 & 0.2336 & 0.0306 & 0.9251 & 1.7692 & 0.2855 & 0.0000 & 0.9996 \\
R10 & 4.3069 & 0.6226 & 0.0056 & 0.9553 & 4.2600 & 0.6260 & 0.0050 & 0.9580 & 3.1889 & 0.7378 & 0.0078 & 0.9646 & 4.5119 & 0.5997 & 0.0025 & 0.9671 \\
R11 & 2.0828 & 0.2755 & 0.0402 & 0.9767 & 2.1050 & 0.2800 & 0.0440 & 0.9750 & 2.0870 & 0.2891 & 0.0634 & 0.9486 & 2.0636 & 0.2818 & 0.0264 & 0.9999 \\
R12 & 2.5539 & -0.1801 & 0.0302 & 0.9850 & 2.5670 & -0.1770 & 0.0340 & 0.9850 & 2.4036 & -0.2148 & 0.0447 & 0.9738 & 2.4438 & -0.1823 & 0.0007 & 0.9999 \\
R13 & 2.1091 & 0.4417 & 0.0264 & 0.9765 & 2.1090 & 0.4390 & 0.0260 & 0.9750 & 2.0391 & 0.4755 & 0.0425 & 0.9551 & 2.0763 & 0.4299 & 0.0090 & 0.9999 \\
R14 & 2.5395 & 0.0800 & 0.0272 & 0.9895 & 2.5450 & 0.0900 & 0.0330 & 0.9900 & 2.3380 & 0.0851 & 0.0424 & 0.9787 & 2.5481 & 0.0784 & 0.0126 & 1.0000 \\
R15 & 3.4591 & 0.5558 & 0.0294 & 0.9791 & 3.4390 & 0.5570 & 0.0290 & 0.9800 & 2.8142 & 0.6201 & 0.0286 & 0.9705 & 3.5276 & 0.5506 & 0.0256 & 0.9997 \\
R16 & 3.6603 & -0.2655 & 0.0411 & 0.9952 & 3.6140 & -0.2740 & 0.0360 & 0.9950 & 2.9813 & -0.3418 & 0.0352 & 0.9909 & 3.5878 & -0.2743 & 0.0109 & 1.0000 \\
R17 & 3.7246 & 1.0666 & 0.0022 & 0.9587 & 3.7180 & 1.0670 & 0.0020 & 0.9600 & 3.0220 & 1.2126 & 0.0045 & 0.9459 & 3.7827 & 1.0479 & 0.0000 & 0.9999 \\
R18 & 1.8799 & -0.1658 & 0.0148 & 0.9714 & 1.8750 & -0.1550 & 0.0190 & 0.9740 & 1.8947 & -0.2130 & 0.0384 & 0.9409 & 1.8457 & -0.1229 & 0.0000 & 0.9997 \\
R19 & 3.6103 & 0.6655 & 0.0053 & 0.9879 & 3.6140 & 0.6680 & 0.0050 & 0.9880 & 2.9708 & 0.7532 & 0.0088 & 0.9751 & 3.7358 & 0.6402 & 0.0005 & 1.0000 \\
R20 & 3.4942 & 0.5061 & 0.0131 & 0.9915 & 3.4810 & 0.5060 & 0.0120 & 0.9910 & 2.9046 & 0.5631 & 0.0158 & 0.9814 & 3.5688 & 0.4814 & 0.0050 & 1.0000 \\
R21 & 1.4567 & 0.6611 & 0.0406 & 0.9339 & 1.4980 & 0.6790 & 0.0510 & 0.9330 & 1.6183 & 0.6959 & 0.0778 & 0.8849 & 1.4010 & 0.7346 & 0.0295 & 0.9996 \\
R22 & 3.9304 & -0.0354 & 0.0073 & 0.9934 & 3.9200 & -0.0340 & 0.0070 & 0.9940 & 3.1821 & -0.0618 & 0.0129 & 0.9879 & 4.0752 & -0.0251 & 0.0000 & 1.0000 \\
R23 & 2.4042 & -0.1042 & 0.0365 & 0.9796 & 2.3930 & -0.1160 & 0.0320 & 0.9770 & 2.2506 & -0.1432 & 0.0460 & 0.9662 & 2.3141 & -0.1220 & 0.0024 & 0.9916 \\
R24 & 2.9458 & -0.0212 & 0.0196 & 0.9866 & 2.9340 & -0.0220 & 0.0190 & 0.9870 & 2.6279 & -0.0467 & 0.0271 & 0.9757 & 2.8793 & -0.0197 & 0.0002 & 1.0000 \\
R25 & 1.9217 & -0.5193 & 0.1525 & 0.9847 & 1.9150 & -0.5130 & 0.1550 & 0.9870 & 1.9181 & -0.5814 & 0.1617 & 0.9732 & 1.8225 & -0.5488 & 0.0966 & 0.9999 \\
R26 & 2.7870 & -0.2842 & 0.0143 & 0.9876 & 2.7750 & -0.2800 & 0.0170 & 0.9880 & 2.5492 & -0.3323 & 0.0280 & 0.9778 & 2.7764 & -0.2580 & 0.0000 & 1.0000 \\
R27 & 3.4177 & -0.0430 & 0.0095 & 0.9922 & 3.3910 & -0.0440 & 0.0090 & 0.9920 & 2.9143 & -0.0659 & 0.0173 & 0.9856 & 3.4873 & -0.0329 & 0.0000 & 1.0000 \\
R28 & 3.4388 & -0.3033 & 0.0233 & 0.9835 & 3.4450 & -0.2990 & 0.0240 & 0.9830 & 2.9204 & -0.3652 & 0.0285 & 0.9795 & 3.3851 & -0.2953 & 0.0002 & 0.9905 \\
R29 & 3.8608 & 0.3331 & 0.0037 & 0.9684 & 3.8650 & 0.3330 & 0.0040 & 0.9670 & 3.0813 & 0.3787 & 0.0075 & 0.9674 & 3.9442 & 0.3412 & 0.0000 & 0.9874 \\
\hline
\end{tabular}
\captionsetup{width=0.7\linewidth}\caption{Comparison of estimated item parameters for the Anxiety dataset using four methods: (1) the 4PL FA model estimated via Bayesian methods in \texttt{JAGS} and (2) \texttt{pymc}, and (3) 4PL IRT model with typical IRT parameterization via Bayesian methods in \texttt{JAGS}, and (4) the 4PL IRT model estimated via the EM algorithm in the \texttt{mirt} package.}
\label{tab2}
\end{sidewaystable}

\begin{sidewaystable}[htbp]
\centering
\begin{tabular}{rrrrrrrrrrrrrrrrrrrr|rr}
\hline
I49 & I27 & I41 & I7 & I38 & I28 & I9 & I47 & I75 & I17 & I76 & I10 & I64 & I45 & I24 & I1 & I68 & I61 & I25 & I2 & $\overline{Y}$ & $\overline{Z}$ \\
\hline
1 & 0 & 0 & 0 & 0 & 0 & 0 & 1 & 0 & 0 & 0 & 0 & 0 & 0 & 0 & 0 & 0 & 0 & 0 & 0 & 2 & 0.52 \\
0 & 0 & 0 & 0 & 0 & 0 & 0 & 1 & 0 & 0 & 0 & 0 & 1 & 0 & 0 & 0 & 0 & 0 & 0 & 0 & 2 & 0.62 \\
0 & 0 & 0 & 0 & 0 & 0 & 0 & 0 & 0 & 0 & 0 & 0 & 1 & 0 & 0 & 0 & 0 & 0 & 1 & 0 & 2 & 0.58 \\
1 & 0 & 0 & 1 & 0 & 0 & 0 & 1 & 0 & 0 & 0 & 0 & 0 & 0 & 0 & 0 & 0 & 0 & 0 & 0 & 3 & 1.36 \\
0 & 0 & 0 & 0 & 1 & 0 & 1 & 1 & 0 & 0 & 0 & 0 & 0 & 0 & 0 & 0 & 0 & 0 & 0 & 0 & 3 & 1.52 \\
1 & 0 & 0 & 0 & 0 & 0 & 0 & 1 & 0 & 0 & 1 & 0 & 0 & 0 & 0 & 0 & 0 & 0 & 0 & 0 & 3 & 0.57 \\
1 & 1 & 0 & 0 & 0 & 0 & 0 & 0 & 0 & 0 & 0 & 0 & 0 & 0 & 1 & 0 & 0 & 0 & 0 & 0 & 3 & 1.27 \\
1 & 0 & 0 & 0 & 0 & 0 & 0 & 1 & 0 & 0 & 0 & 0 & 0 & 0 & 0 & 1 & 0 & 0 & 0 & 0 & 3 & 0.61 \\
0 & 0 & 0 & 0 & 0 & 0 & 0 & 0 & 0 & 0 & 1 & 0 & 0 & 1 & 0 & 1 & 0 & 0 & 0 & 0 & 3 & 0.67 \\
1 & 0 & 0 & 0 & 1 & 0 & 0 & 0 & 0 & 0 & 0 & 0 & 0 & 0 & 0 & 0 & 1 & 0 & 0 & 0 & 3 & 1.19 \\
0 & 0 & 0 & 0 & 0 & 0 & 0 & 1 & 0 & 0 & 0 & 0 & 0 & 1 & 0 & 0 & 1 & 0 & 0 & 0 & 3 & 1.91 \\
0 & 0 & 0 & 0 & 1 & 0 & 1 & 0 & 0 & 0 & 0 & 0 & 0 & 0 & 0 & 0 & 0 & 1 & 0 & 0 & 3 & 1.73 \\
1 & 0 & 0 & 0 & 0 & 0 & 0 & 0 & 1 & 0 & 0 & 0 & 0 & 0 & 0 & 0 & 0 & 1 & 0 & 0 & 3 & 0.69 \\
1 & 0 & 0 & 0 & 0 & 0 & 0 & 1 & 0 & 0 & 0 & 0 & 0 & 0 & 0 & 0 & 0 & 0 & 1 & 0 & 3 & 0.94 \\
0 & 0 & 0 & 0 & 1 & 0 & 0 & 1 & 0 & 0 & 0 & 0 & 0 & 0 & 0 & 0 & 0 & 0 & 1 & 0 & 3 & 1.18 \\
1 & 0 & 0 & 0 & 0 & 0 & 0 & 0 & 0 & 0 & 0 & 1 & 0 & 0 & 0 & 0 & 0 & 0 & 1 & 0 & 3 & 1.02 \\
1 & 0 & 0 & 0 & 0 & 0 & 0 & 0 & 0 & 0 & 0 & 0 & 1 & 0 & 0 & 0 & 0 & 0 & 1 & 0 & 3 & 0.65 \\
0 & 0 & 0 & 0 & 0 & 0 & 1 & 1 & 0 & 0 & 0 & 0 & 0 & 0 & 0 & 0 & 0 & 0 & 0 & 1 & 3 & 1.15 \\
0 & 0 & 0 & 0 & 0 & 0 & 1 & 0 & 0 & 0 & 0 & 0 & 0 & 0 & 0 & 0 & 0 & 0 & 1 & 1 & 3 & 1.11 \\
0 & 0 & 0 & 0 & 0 & 0 & 0 & 0 & 0 & 0 & 0 & 0 & 0 & 1 & 0 & 0 & 0 & 0 & 1 & 1 & 3 & 0.95 \\
1 & 0 & 0 & 0 & 0 & 1 & 1 & 0 & 1 & 0 & 0 & 0 & 0 & 0 & 0 & 0 & 0 & 0 & 0 & 0 & 4 & 0.86 \\
1 & 0 & 0 & 0 & 0 & 0 & 1 & 1 & 1 & 0 & 0 & 0 & 0 & 0 & 0 & 0 & 0 & 0 & 0 & 0 & 4 & 1.35 \\
0 & 0 & 0 & 0 & 0 & 1 & 0 & 1 & 1 & 0 & 1 & 0 & 0 & 0 & 0 & 0 & 0 & 0 & 0 & 0 & 4 & 0.59 \\
1 & 0 & 0 & 0 & 0 & 0 & 0 & 1 & 0 & 0 & 1 & 1 & 0 & 0 & 0 & 0 & 0 & 0 & 0 & 0 & 4 & 1.15 \\
1 & 0 & 0 & 0 & 0 & 0 & 0 & 0 & 1 & 0 & 1 & 1 & 0 & 0 & 0 & 0 & 0 & 0 & 0 & 0 & 4 & 0.73 \\
1 & 0 & 1 & 0 & 0 & 0 & 0 & 1 & 0 & 0 & 0 & 0 & 1 & 0 & 0 & 0 & 0 & 0 & 0 & 0 & 4 & 0.69 \\
1 & 0 & 0 & 0 & 0 & 0 & 0 & 1 & 1 & 0 & 0 & 0 & 1 & 0 & 0 & 0 & 0 & 0 & 0 & 0 & 4 & 0.76 \\
0 & 0 & 1 & 0 & 0 & 0 & 0 & 1 & 0 & 0 & 1 & 0 & 1 & 0 & 0 & 0 & 0 & 0 & 0 & 0 & 4 & 0.65 \\
0 & 0 & 0 & 0 & 0 & 1 & 1 & 1 & 0 & 0 & 0 & 0 & 0 & 0 & 0 & 1 & 0 & 0 & 0 & 0 & 4 & 1.26 \\
1 & 1 & 0 & 0 & 0 & 0 & 0 & 0 & 0 & 0 & 0 & 1 & 0 & 0 & 0 & 1 & 0 & 0 & 0 & 0 & 4 & 1.12 \\
\hline
\end{tabular}
\captionsetup{width=0.7\linewidth}
\caption{Item-level response patterns, total scores $\overline{Y}$, and averaged non-guessing and not-inattention (NGNI) total scores $\overline{Z}$ for unique response patterns in the MSATB dataset, computed across two \texttt{JAGS} runs. Only the first 30 patterns are shown.}
\label{tabNGNI3}
\end{sidewaystable}

\newpage
\begin{table}[ht]
\centering
\begin{tabular}{r|rrrr}
\hline
Parameter & Discrimination & Difficulty & Guessing & Inattention\\
\hline
JAGS & 0.0779 & 0.0202 & 0.0019 & 0.0036 \\
pymc & 0.0891 & 0.0240 & 0.0028 & 0.0030 \\
\hline
\end{tabular}
\caption{Mean squared error comparison of 4FA parameter estimates from \texttt{JAGS} and \texttt{pymc} computed with respect to \texttt{mirt} estimates, based on the MSATB dataset.}
\label{mse_mirt_tab_1}
\end{table}

\begin{table}[ht]
\centering
\begin{tabular}{r|rrrr}
\hline
Parameter & Discrimination & Difficulty & Guessing & Inattention\\
\hline
JAGS & 0.0472 & 0.0006 & 0.0003 & 0.0005 \\
pymc & 0.0687 & 0.0005 & 0.0003 & 0.0005 \\
\hline
\end{tabular}
\caption{Mean squared error comparison of 4FA parameter estimates from \texttt{JAGS} and \texttt{pymc} computed with respect to \texttt{mirt} estimates, based on the Anxiety dataset.}
\label{mse_mirt_tab_2}
\end{table}

\end{document}